\pdfoutput=1
\RequirePackage{ifpdf}
\ifpdf 
\documentclass[pdftex]{sigma}
\else
\documentclass{sigma}
\fi

\usepackage[all]{xy}
\usepackage{enumerate}

\numberwithin{equation}{section}

\newtheorem{Theorem}{Theorem}[section]
\newtheorem{Corollary}[Theorem]{Corollary}
\newtheorem{Lemma}[Theorem]{Lemma}
\newtheorem{Proposition}[Theorem]{Proposition}
{\theoremstyle{definition}
\newtheorem{Definition}[Theorem]{Definition}
\newtheorem{Remark}[Theorem]{Remark}
}

\DeclareMathOperator{\Sc}{Sc}
\DeclareMathOperator{\Ric}{Ric}
\DeclareMathOperator{\Tr}{Tr}

\begin{document}

\allowdisplaybreaks

\renewcommand{\PaperNumber}{016}

\FirstPageHeading

\ShortArticleName{Second Order Symmetries of the Conformal Laplacian}

\ArticleName{Second Order Symmetries of the Conformal Laplacian}

\Author{Jean-Philippe MICHEL~$^\dag$, Fabian RADOUX~$^\dag$ and Josef \v{S}ILHAN~$^\ddag$}

\AuthorNameForHeading{J.-P.~Michel, F.~Radoux and J.~\v{S}ilhan}

\Address{$^\dag$~Department of Mathematics of the University of Li\`ege,\\
\hphantom{$^\dag$}~Grande Traverse 12, 4000 Li\`ege, Belgium}
\EmailD{\href{mailto:jean-philippe.michel@ulg.ac.be}{jean-philippe.michel@ulg.ac.be},
\href{mailto:fabian.radoux@ulg.ac.be}{fabian.radoux@ulg.ac.be}}
\URLaddressD{\url{http://www.geodiff.ulg.ac.be/recherche/fradoux/},\\
\hspace*{13.5mm}\url{http://www.geodiff.ulg.ac.be/recherche/michel/}}

\Address{$^\ddag$~Department of Algebra and Geometry of the Masaryk University in Brno,\\
\hphantom{$^\ddag$}~Jan\`a\v{c}kovo n\`am.~2a, 662 95 Brno, Czech Republic}
\EmailD{\href{mailto:silhan@math.muni.cz}{silhan@math.muni.cz}}
\URLaddressD{\url{http://www.math.muni.cz/~silhan/}}

\ArticleDates{Received October 25, 2013, in f\/inal form February 05, 2014; Published online February 14, 2014}

\Abstract{Let $(M,{\rm g})$ be an arbitrary pseudo-Rie\-man\-nian manifold of dimension at least~$3$.
We determine the form of all the conformal symmetries of the conformal (or Yamabe) Laplacian on $(M,{\rm g})$, which are given by dif\/ferential operators of second order.
They are constructed from conformal Killing $2$-tensors satisfying a~natural and conformally invariant
condition.
As a~consequence, we get also the classif\/ication of the second order symmetries of the conformal
Laplacian.
Our results generalize the ones of Eastwood and Carter, which hold on conformally f\/lat and Einstein
manifolds respectively.
We illustrate our results on two families of examples in dimension three.}


\Classification{58J10; 53A30; 70S10; 53D20; 53D55}

\section{Introduction}

We work over a~pseudo-Rie\-man\-nian manifold $(M,{\rm g})$ of dimension $n\geq 3$, with Levi-Civita connection
$\nabla$ and scalar curvature ${\rm Sc}$.
Our main result is the classif\/ication of all the dif\/ferential operators $D_1$ of second order such that
the relation
\begin{gather}
\label{SymConf}
\Delta_{Y}D_1=D_2\Delta_{Y}
\end{gather}
holds for some dif\/ferential operator $D_2$, where $\Delta_{Y}:=\nabla_a{\rm
g}^{ab}\nabla_b-\frac{n-2}{4(n-1)}{\rm Sc}$ is the Yamabe Laplacian.
Such operators $D_1$ are called conformal symmetries of order $2$ of $\Delta_{Y}$.
They preserve the kernel of $\Delta_{Y}$, i.e.\ the solution space of the equation $\Delta_{Y}\psi=0$,
$\psi\in\mathcal{C}^{\infty}(M)$.
Under a~conformal change of metric, $\hat{{\rm g}}=e^{2\Upsilon}{\rm g}$, $\Upsilon
\in\mathcal{C}^{\infty}(M)$, the Yamabe Laplacian transforms as
\begin{gather*}
\widehat{\Delta_{Y}}=e^{-\frac{n+2}{2}\Upsilon}\circ\Delta_{Y}\circ e^{\frac{n-2}{2}\Upsilon},
\end{gather*}
so that each conformal symmetry $D_1$ of $\Delta_{Y}$ gives rise to one of $\widehat{\Delta_{Y}}$ given by
\begin{gather*}
\widehat{D_1}=e^{-\frac{n-2}{2}\Upsilon}\circ D_1\circ e^{\frac{n-2}{2}\Upsilon}.
\end{gather*}
This emphasizes the conformal nature of the problem and justify our choice of the Yamabe Laplacian, rather
than the more usual Laplace--Beltrami one, $\Delta:=\nabla_a{\rm g}^{ab}\nabla_b$.
Over f\/lat pseudo-Euclidean space, the classif\/ication of conformal symmetries up to second order is due
to Boyer, Kalnins and Miller~\cite{BKM76}, who use it to study the $R$-separation of variables of the
Laplace equation $\Delta \Psi=0$.
More generally, Kalnins and Miller provide an intrinsic characterization for $R$-separation of this
equation on $(M,{\rm g})$ in terms of second order conformal symmetries~\cite{KMi82}.
Thus, classifying those symmetries happens to be a~basic problem in the theory of separation of variables.
A new input into the quest of conformal symmetries has been given by the work of Eastwood~\cite{Eas05}.
He classif\/ied indeed the conformal symmetries of any order over the conformally f\/lat space and
exhibited their interesting algebraic structure.
This leads to a~number of subsequent works, dealing with other invariant operators~\cite{ELe08,GSi09,Vla12}.

Using principal symbol maps, one can extract two informations from the equation~\eqref{SymConf}: the
operators $D_1$ and $D_2$ have the same principal symbol and the latter is a~conformal Killing $2$-tensor,
i.e.\ a constant of motion of the geodesic f\/low, restricted to the null cone.
One looks then for a~right inverse to the principal symbol maps, called a~quantization map, which
associates with each conformal Killing tensor a~conformal symmetry of $\Delta_{Y}$.
For Killing vector f\/ields this is trivial.
If $K$ is a~$2$-tensor, Carter proves that if the minimal prescription
\begin{gather*}
K\mapsto\nabla_aK^{ab}\nabla_b
\end{gather*}
satisf\/ies $[\Delta_{Y},\nabla_aK^{ab}\nabla_b]=0$, then $K$ is Killing.
Moreover, he shows that if $(M,{\rm g})$ is Einstein, i.e.\ if $\Ric=\frac{1}{n}{\rm Sc}{\rm g}$
with $\Ric$ the Ricci tensor, the fact that $K$ is Killing is suf\/f\/icient to ensure that the
minimal prescription above is a~symmetry of $\Delta_{Y}$ (for application to the separation of variables,
see~\cite{BCR02'}).
Besides, in~\cite{Eas05}, Eastwood def\/ines conformally invariant operators on an arbitrary
pseudo-Rie\-man\-nian manifold, which coincide with the conformal symmetries of $\Delta_{Y}$ on the f\/lat
space.
These operators are given by means of the natural and conformally invariant quantization
$\mathcal{Q}_{\lambda_{0},\lambda_{0}}$ (where $\lambda_{0}=\frac {n-2}{2n}$), developed
in~\cite{CSi09, MRa09,Rad09,Sil09}.
Explicitly, if $X$ is a~vector f\/ield and $K$ a~symmetric trace-less $2$-tensor,
$\mathcal{Q}_{\lambda_{0},\lambda_{0}}(X)$ and $\mathcal{Q}_{\lambda_{0},\lambda_{0}}(K)$ are
dif\/ferential operators acting between $\lambda_{0}$-densities def\/ined in the following way:
\begin{gather*}
\mathcal{Q}_{\lambda_{0},\lambda_{0}}(X)=X^a\nabla_a+\frac{n-2}{2n}(\nabla_a X^a),
\\
\mathcal{Q}_{\lambda_{0},\lambda_{0}}(K)=K^{ab}\nabla_a\nabla_b+\frac{n}{n\!+\!2}
\left((\nabla_aK^{ab})\nabla_b+\frac{n\!-\!2}{4(n\!+\!1)}(\nabla_a\nabla_b K^{ab})\right)-\frac{n\!+\!2}{4(n\!+\!1)}\Ric_{ab}K^{ab}.
\end{gather*}
In the conformally f\/lat case, all the conformal symmetries of second order are of the type
$\mathcal{Q}_{\lambda_{0},\lambda_{0}}(K+X+c)$, where $c\in\mathbb{R}$, $X$ is a~conformal Killing vector
f\/ield and $K$ is a~conformal Killing $2$-tensor.
Thanks to the conformal covariance of $\Delta_{Y}$ one can show that
$\mathcal{Q}_{\lambda_{0},\lambda_{0}}(X)$ is still a~conformal symmetry of $\Delta_{Y}$ on an arbitrary
pseudo-Rie\-man\-nian manifold, if $X$ is a~conformal Killing vector f\/ield.
However, as pointed out by Eastwood in~\cite{Eas05}, it is unclear whether
$\mathcal{Q}_{\lambda_{0},\lambda_{0}}(K)$ is a~conformal symmetry when $K$ is a~conformal Killing $2$-tensor.

Our strategy relies on the properties of the quantization map $\mathcal{Q}_{\lambda_{0},\lambda_{0}}$ and
on the classif\/ication of natural and conformally invariant operators acting on prescribed subspaces of
symbols.
This method has been developed f\/irst on conformally f\/lat manifolds, in~\cite{Mic11b}.
In that case, the map~$\mathcal{Q}_{\lambda_{0},\lambda_{0}}$ is a~conformally equivariant
quantization~\cite{DLO99}, and the author proved that it is precisely the bijective map between conformal
Killing tensors and conformal symmetries of~$\Delta_{Y}$, discovered by Eastwood.
The description of conformal symmetries on arbitrary pseudo-Rie\-man\-nian manifolds is more involved, even at
order $2$.
Namely, there exists a~conformal symmetry with principal symbol~$K$ if and only if~$K$ is a~conformal
Killing tensor and~$\mathbf{Obs}(K)^\flat$ is an exact one-form.
Here, $\mathbf{Obs}$ is a~natural and conformally invariant operator which reads, in abstract index notation, as
\begin{gather*}
\mathbf{Obs}(K)^a=\frac{2(n-2)}{3(n+1)}\left(\mathrm{C}^r{}_{st}{}^{a}\nabla_r-3\mathrm{A}_{st}{}^a\right)K^{st},
\end{gather*}
where $\mathrm{C}$ denotes the Weyl tensor and $\mathrm{A}$ the Cotton--York tensor.
If $\mathbf{Obs}(K)^{\flat}$ is equal to the exact one-form $-2 df$, with $f\in\mathcal{C}^{\infty}(M)$,
then the operators
\begin{gather*}
\mathcal{Q}_{\lambda_{0},\lambda_{0}}(K+X+c)+f
\end{gather*}
are conformal symmetries of $\Delta_{Y}$ for all conformal Killing vector f\/ield $X$ and constant
$c\in\mathbb{R}$.
As a~consequence, $\mathcal{Q}_{\lambda_{0},\lambda_{0}}(K)$ is a~conformal symmetry of $\Delta_{Y}$ if and
only if $\mathbf{Obs}(K)=0$.

We illustrate our results on two examples in dimension three.
In the f\/irst one, the space~$\mathbb{R}^3$ is endowed with the most general Riemannian metric admitting
a~Killing $2$-tensor $K$, which is diagonal in orthogonal coordinates~\cite{Per90}.
Then, $\mathbf{Obs}(K)^\flat$ is a~non-trivial exact $1$-form and, up to our knowledge, the symmetry of
$\Delta_{Y}$ that we obtain is new.
In the second one, we consider a~conformal St\"ackel metric ${\rm g}$ on $\mathbb{R}^3$ with one ignorable
coordinate.
Such a~metric admits an irreducible conformal Killing tensor $K$.
Using the generic form of ${\rm g}$ and $K$ given in the reference~\cite{BKM78}, we obtain that
$\mathbf{Obs}(K)^\flat$ is a~non-exact $1$-form in general.
This means there are no conformal symmetries of $\Delta_{Y}$ with principal symbol $K$ in general.

We detail now the content of the paper.

In Section~\ref{Section2},
we introduce the basic spaces: the one of tensor densities $\mathcal{F}_\lambda(M)$
of weight $\lambda\in\mathbb{R}$, the one of dif\/ferential operators $\mathcal{D}_{\lambda,\mu}(M)$ acting
between $\lambda$- and $\mu$-densities, the one of symbols $\mathcal{S}_\delta(M)$ with
$\delta=\mu-\lambda$.
Then, we def\/ine the Yamabe Laplacian $\Delta_{Y}$ as an element of
$\mathcal{D}_{\lambda_{0},\mu_{0}}(M)$, with $\lambda_{0}=\frac{n-2}{2n}$ and $\mu_{0}=\frac{n+2}{2n}$, so
that it becomes a~conformally invariant operator.
Finally we introduce our main tool, namely the natural and conformally invariant quantization
\begin{gather*}
\mathcal{Q}_{\lambda,\mu}:
\
\mathcal{S}_{\mu-\lambda}(M)
\
\rightarrow
\
\mathcal{D}_{\lambda,\mu}(M),
\end{gather*}
and we provide explicit formulas for it.

In Section~\ref{Section3},
we classify the natural and conformally invariant operators between some subspaces of symbols.
Among the operators we obtain (and which are crucial for understanding of 2nd order symmetries), one of
them, $\mathbf{G}$, is classical, whereas another one, $\mathbf{Obs}$, acting on symbols of second degree,
is new and admits no counterpart on f\/lat space.
We obtain also an analogous classif\/ication for higher order trace-free symbols where the situation is
much more complicated.
Note that the discovered operators act between source and target spaces of well-known conformally invariant
operators, which appear in the generalized BGG sequence~\cite{CSS01}.
It would be interesting to understand better the relations between all these conformal operators.

In Section~\ref{sym2nd} lies our main result.
After def\/ining the spaces of conformal symmetries and of conformal Killing tensors, we prove that, on
symbols $K$ of degree $2$, we have
\begin{gather*}
(\mathcal{Q}_{\lambda_{0},\mu_{0}})^{-1}
\left(\Delta_{Y}\mathcal{Q}_{\lambda_{0},\lambda_{0}}(K)-\mathcal{Q}_{\mu_{0},\mu_{0}}(K)\Delta_{Y}\right)
=2\mathbf{G}(K)+\mathbf{Obs}(K).
\end{gather*}
The kernel of $\mathbf{G}$ is precisely the space of conformal Killing tensors, whereas $\mathbf{Obs}(K)$
is the obstruction for a~conformal Killing $2$-tensor to provide a~conformal symmetry of the form
$\mathcal{Q}_{\lambda_{0},\lambda_{0}}(K)$.
The full description of conformal symmetries of 2nd order of $\Delta_{Y}$ easily follows.
Using that $\mathcal{Q}_{\lambda_{0},\lambda_{0}}(K)=\mathcal{Q}_{\mu_{0},\mu_{0}}(K)$ for Killing
$2$-tensors, we deduce also the classif\/ication of second order symmetries of $\Delta_{Y}$, which satisfy
by def\/inition $[\Delta_{Y},D_1]=0$.

In Section~\ref{Section5}, we provide two examples illustrating our main result.
In the f\/irst one, the Killing tensor $K$ is such that $\mathbf{Obs}(K)^\flat$ is a non-vanishing but exact one-form.
In the second example, we provide several conformal Killing tensors $K$ such that $\mathbf{Obs}(K)^\flat$
is a non-exact one-form.
Hence, there is no conformal symmetry with such $K$ as principal symbols.

\section{Conformal geometry, dif\/ferential operators,\\ and their symbols}\label{Section2}

Throughout this paper, we employ the abstract index notation from~\cite{PRi84}.
That is, on a~smooth manifold~$M$, $v^a$ denotes a~section of the tangent bundle $TM$, $v_a$ a~section of
the cotangent bundle $T^*M$ and e.g.\ $v^{ab}{}_c$ a~section of $TM \otimes TM \otimes T^*M$.
The letters $a$, $b$, $c$, $d$ and $r$, $s$, $t$ are reserved for abstract indices.
Repetition of an abstract index in the covariant and contravariant position means contraction, e.g.\
$v^{ab}_b$ is a~section of $TM$.
In few places we use concrete indices attached to a~coordinate system.
This is always explicitly stated and we denote such indices by letters $i$, $j$, $k$, $l$ to avoid confusion with
abstract indices.
We always use the Einstein's summation convention for indices, except if stated otherwise.

\subsection{Basic objects}\label{basic}

Let $M$ be a~$n$-dimensional smooth manifold.
If $\lambda\in\mathbb{R}$, the vector bundle of {\it $\lambda$-densities}, \mbox{$F_{\lambda}(M)\to M$}, is a~line
bundle associated with $P^{1}M$, the linear frame bundle over $M$:
\begin{gather*}
F_{\lambda}(M)=P^1M\times_{\rho}\mathbb{R},
\end{gather*}
where the representation $\rho$ of the group ${\rm GL}(n,\mathbb{R})$ on $\mathbb{R}$ is given by
\begin{gather*}
\rho(A)e=\vert\det A\vert^{-\lambda}e,
\qquad
\forall\, A\in {\rm GL}(n,\mathbb{R}),
\quad
\forall\, e\in\mathbb{R}.
\end{gather*}
We denote by $\mathcal{F}_{\lambda}(M)$ the space of smooth sections of this bundle.
Since $F_{\lambda}(M)$ is associated with $P^1M$, the space $\mathcal{F}_{\lambda}(M)$ is endowed with
canonical actions of $\mathrm{Dif\/f}(M)$ and $\mathrm{Vect}(M)$.
If $(x^{1},\ldots,x^{n})$ is a~coordinate system on $M$, we denote by $|Dx|^{\lambda}$ the local
$\lambda$-density equal to $[({\rm Id},1)]$, where ${\rm Id}$ is the identity frame in the coordinates
system $(x^{1},\ldots,x^{n})$.

Actually, a~$\lambda$-density $\varphi$ at a~point $x\in M$ can be viewed as a~map on $\wedge^{n}T_{x}M$
with values in~$\mathbb{R}$ such that
\begin{gather*}
\varphi(c X_{1}\wedge\cdots\wedge X_{n})=|c|^{\lambda}\varphi(X_{1}\wedge\cdots\wedge X_{n})
\end{gather*}
for all $X_{1},\ldots,X_{n}\in T_{x}M$ and $c\in\mathbb{R}$.
The $\lambda$-density $|Dx|^{\lambda}$ is then the $\lambda$-density equal to one on
$\partial_{1}\wedge\cdots\wedge\partial_{n}$, where $\partial_{1},\ldots,\partial_{n}$ denotes the
canonical basis of $T_{x}M$ corresponding to the coordinate system $(x^{1},\ldots,x^{n})$.

If a~$\lambda$-density $\varphi$ reads locally $f|Dx|^{\lambda}$, where $f$ is a~local function, then the
Lie derivative of $\varphi$ in the direction of a~vector f\/ield $X$ reads locally
\begin{gather}
\label{LieDerivative}
L_{X}^{\lambda}\varphi=\big(X.f+\lambda\big(\partial_{i}X^{i}\big)f\big)|Dx|^{\lambda}.
\end{gather}
It is possible to def\/ine the multiplication of two densities.
If $\varphi_{1}$ reads locally $f|Dx|^{\lambda}$ and if $\varphi_{2}$ reads locally $g|Dx|^{\delta}$, then
$\varphi_{1}\varphi_{2}$ reads locally $fg|Dx|^{\lambda+\delta}$.

On a~pseudo-Rie\-man\-nian manifold $(M,{\rm g})$, it is possible to def\/ine in a~natural way
a~$\lambda$-density.
In a~coordinate system, this $\lambda$-density reads
\begin{gather*}
|\mathrm{Vol}_{\rm g}|^{\lambda}=|\det{\rm g}|^{\frac{\lambda}{2}}|Dx|^{\lambda},
\end{gather*}
where $|\det{\rm g}|$ denotes the absolute value of the determinant of the matrix representation of ${\rm
g}$ in the coordinate system.

We shall denote by $\mathcal{D}_{\lambda,\mu}(M)$ the space of dif\/ferential operators from
$\mathcal{F}_{\lambda}(M)$ to $\mathcal{F}_{\mu}(M)$.
It is the space of linear maps between $\mathcal{F}_{\lambda}(M)$ and $\mathcal{F}_{\mu}(M)$ that read in
trivialization charts as dif\/ferential operators.
The actions of $\mathrm{Vect}(M)$ and $\mathrm{Dif\/f}(M)$ on $\mathcal{D}_{\lambda,\mu}(M)$ are induced by
the actions on tensor densities: if $\mathcal{L}_{X}D$ denotes the Lie derivative of the dif\/ferential
operator $D$ in the direction of the vector f\/ield $X$, we have
\begin{gather*}
\mathcal{L}_{X}D=L^{\mu}_{X}\circ D-D\circ L_{X}^{\lambda},
\qquad
\forall\, D\in\mathcal{D}_{\lambda,\mu}(M)
\qquad
\text{and}
\qquad
\forall\, X\in\mathrm{Vect}(M).
\\
\phi\cdot D=\phi\circ D\circ\phi^{-1},
\qquad
\forall\, D\in\mathcal{D}_{\lambda,\mu}(M)
\qquad
\text{and}
\qquad
\forall\, \phi\in\mathrm{Diff}(M).
\end{gather*}

The space $\mathcal{D}_{\lambda,\mu}(M)$ is f\/iltered by the order of dif\/ferential operators.
We denote by $\mathcal{D}^k_{\lambda,\mu}(M)$ the space of dif\/ferential operators of order $k$.
It is well-known that this f\/iltration is preserved by the action of local dif\/feomorphisms.

{\sloppy On a~pseudo-Rie\-man\-nian manifold $(M,{\rm g})$, it is easy to build an isomorphism between
$\mathcal{D}_{\lambda,\mu}(M)$ and $\mathcal{D}(M)$, the space of dif\/ferential operators acting between
functions.
Indeed, thanks to the canonical densities built from $|\mathrm{Vol}_{\rm g}|$, all operators
$D\in\mathcal{D}_{\lambda,\mu}(M)$ can be pulled-back on functions as follows
\begin{gather}
\begin{split}
& \xymatrix{
\mathcal{F}_\lambda(M)\ar[rrr]^{D} &&& \mathcal{F}_\mu(M) \\
\mathcal{C}^{\infty}(M) \ar[u]^{|\mathrm{Vol}_{\rm g}|^\lambda}\ar[rrr]_{|\mathrm{Vol}_{\rm g}|^{-\mu}\circ D\circ|\mathrm{Vol}_{\rm g}|^{\lambda}}
&&& \mathcal{C}^{\infty}(M)\ar[u]_{|\mathrm{Vol}_{\rm g}|^\mu}
}\end{split}\label{IsoDlm-D}
\end{gather}

}

The space of \emph{symbols} is the graded space associated with $\mathcal{D}_{\lambda,\mu}(M)$: it is then
equal to
\begin{gather*}
{\mathrm{gr}}\mathcal{D}_{\lambda,\mu}(M):=\bigoplus_{k=0}^\infty\mathcal{D}_{\lambda,\mu}^{k}
(M)/\mathcal{D}_{\lambda,\mu}^{k-1}(M).
\end{gather*}
The canonical projection
$\sigma_{k}:
\mathcal{D}_{\lambda,\mu}^{k}(M)
\rightarrow
\mathcal{D}_{\lambda,\mu}^{k}(M)/\mathcal{D}_{\lambda,\mu}^{k-1}(M)$
is called the principal symbol map.
As the actions of ${\rm Dif\/f}(M)$ and  ${\rm Vect}(M)$ preserve the f\/iltration of
$\mathcal{D}_{\lambda,\mu}(M)$, they induce actions of ${\rm Dif\/f}(M)$ and ${\rm Vect}(M)$ on the space
of symbols.

Let $\delta=\mu-\lambda$ be the shift of weights.
If the sum of the $k$-order terms of $D\in\mathcal{D}^k_{\lambda,\mu}$ in a~coordinate system
$(x^{1},\ldots,x^{n})$ reads
\begin{gather*}
D^{i_1\dots i_{k}}\partial_{i_1}\cdots\partial_{i_k}
\end{gather*}
and if $(x^{i},p_{i})$ is the coordinate system on $T^{*}M$ canonically associated with
$(x^{1},\ldots,x^{n})$, then we get the following identif\/ication:
\begin{gather*}
\sigma_k(D)
\qquad
\longleftrightarrow
\qquad
D^{i_1\dots i_{k}}p_{i_1}\cdots p_{i_k}.
\end{gather*}
Thus, the space of symbols of degree $k$ can be viewed as the space
$\mathcal{S}_{\delta}^k(M):=\mathrm{Pol}^{k}(T^*M)\otimes_{\mathcal{C}^{\infty}(M)}\mathcal{F}_\delta(M)$,
where $\mathrm{Pol}^{k}(T^*M)$ denotes the space of real functions on $T^*M$ which are polynomial functions
of degree $k$ in the f\/ibered coordinates of $T^{*}M$.
The algebra $\mathcal{S}(M):=\mathrm{Pol}(T^*M)$ is clearly isomorphic to the algebra $\Gamma(S TM)$ of
symmetric tensors and depending on the context we will refer to its elements as symbols, functions on
$T^*M$ or symmetric tensors on $M$.

Let us recall that, if $S_{1},S_{2}\in\mathcal{S}(M)$, then the Poisson bracket of $S_1$ and $S_2$, denoted
by $\{S_1,S_2\}$, is def\/ined in a~canonical coordinate system $(x^{i},p_{i})$ of $T^*M$ in the following
way:
\begin{gather}
\label{Poisson}
\{S_{1},S_{2}\}=(\partial_{p_i}S_{1})\big(\partial_{x^{i}}S_{2}\big)-(\partial_{p_i}S_{2})(\partial_{x^{i}}S_{1}).
\end{gather}

We conclude this subsection by two properties of the principal symbol map linked to the composition and to
the commutator of dif\/ferential operators.
For all $k,l\in\mathbb{N}$, we have:
\begin{gather}
\label{SymbolProduct}
\sigma_{k+l}(A\circ B) = \sigma_k(A)\sigma_l(B),
\\
\label{SymbolPoisson}
\sigma_{k+l-1}([A,B]) = \{\sigma_k(A),\sigma_l(B)\},
\end{gather}
where $A$ and $B$ are elements of $\mathcal{D}(M)$ of order $k$ and $l$ respectively.

\subsection{Pseudo-Riemannian and conformal geometry}\label{Par:notation0}

Let $(M,{\rm g})$ be a~pseudo-Rie\-man\-nian manifold.
The isometries $\Phi$ of $(M,{\rm g})$ are the dif\/feo\-morphisms of $M$ that preserve the metric~${\rm g}$,
i.e.\ $\Phi^*{\rm g}={\rm g}$.
Their inf\/initesimal counterparts $X\in\mathrm{Vect}(M)$ are called Killing vector f\/ields, they satisfy
$L_X{\rm g}=0$, with $L_X{\rm g}$ the Lie derivative of ${\rm g}$ along $X$.

Given the Levi-Civita connection $\nabla$ corresponding to the metric ${\rm g}$, the Riemannian curvature
tensor, which reads as $\mathrm{R}_{ab}{}^c{}_d$ in abstract index notation, is given by
$[\nabla_a,\nabla_b]v^c=\mathrm{R}_{ab}{}^c{}_dv^d$ for a~tangent vector f\/ield $v^c$.
Then, one gets the Ricci tensor by taking a~trace of the Riemann tensor, which is indicated by repeated
indices: $\Ric_{bd}=\mathrm{R}_{ab}{}^a{}_d$.
By contraction with the metric, the Ricci tensor leads to the scalar curvature $\Sc={\rm
g}^{ab}\Ric_{ab}$.

A conformal structure on a~smooth manifold $M$ is given by the conformal class $[{\rm g}]$ of
a~pseudo-Rie\-man\-nian metric ${\rm g}$, where two metrics ${\rm g}$ and $\hat{{\rm g}}$ are conformally
related if $\hat{{\rm g}}=e^{2\Upsilon}{\rm g}$, for some function $\Upsilon\in\mathcal{C}^{\infty}(M)$.
The conformal dif\/feomorphisms $\Phi$ of $(M,[{\rm g}])$ are those which preserve the conformal
structure~$[{\rm g}]$, i.e.\ there exists $\Upsilon\in\mathcal{C}^{\infty}(M)$ such that $\Phi^*{\rm
g}=e^{2\Upsilon}{\rm g}$.
Their inf\/initesimal counterparts $X\in\mathrm{Vect}(M)$ are called conformal Killing vector f\/ields,
they satisfy $L_X{\rm g}=f_{X}{\rm g}$, for some function $f_{X}\in\mathcal{C}^{\infty}(M)$.

Let $(x^i,p_i)$ be a~canonical coordinate system on $T^*M$.
If $M$ is endowed with a~metric~${\rm g}$, we def\/ine the metric symbol and the trace operator by,
respectively,
\begin{gather*}
H={\rm g}^{ij}p_ip_j
\qquad
\text{and}
\qquad
\Tr={\rm g}_{ij}\partial_{p_i}\partial_{p_j}.
\end{gather*}
Note that the symbol $|\mathrm{Vol}_{\rm g}|^{2/n}H\in\mathcal{S}_{2/n}$ and the operator
$|\mathrm{Vol}_{\rm g}|^{-2/n}\Tr:
\mathcal{S}_\delta\rightarrow\mathcal{S}_{\delta-2/n}$ are
conformally invariant.
In consequence, we get a~conformally invariant decomposition
\begin{gather}
\label{S_ks}
S^kTM=\bigoplus_{0\leq2s\leq k}S^{k,s}TM,
\end{gather}
where $S\in\mathcal{S}^{k,s}(M):=\Gamma(S^{k,s} TM)$ is of the form $S=H^s S_0$ with $\Tr S_0=0$.

\subsection{The conformal Laplacian}

Starting from a~pseudo-Rie\-man\-nian manifold $(M,{\rm g})$ of dimension $n$, one can def\/ine the Yamabe
Laplacian, acting on functions, in the following way:
\begin{gather*}
\Delta_{Y}:=\nabla_a{\rm g}^{ab}\nabla_b-\frac{n-2}{4(n-1)}\Sc,
\end{gather*}
where $\nabla$ denotes the Levi-Civita connection of ${\rm g}$ and $\Sc$ the scalar curvature.
For the conformally related metric $\hat{{\rm g}}=e^{2\Upsilon}{\rm g}$, the associated Yamabe Laplacian is
given by
\begin{gather*}
\widehat{\Delta_{Y}}=e^{-\frac{n+2}{2}\Upsilon}\circ\Delta_{Y}\circ e^{\frac{n-2}{2}\Upsilon}.
\end{gather*}
According to the transformation law $|\mathrm{Vol}_{\hat{{\rm g}}}|=e^{n\Upsilon}|\mathrm{Vol}_{\rm g}|$
and to the diagram~\eqref{IsoDlm-D}, this translates into the conformal invariance of $\Delta_{Y}$ viewed
as an element of $\mathcal{D}_{\lambda_0,\mu_0}(M)$, for the specif\/ic weights
\begin{gather}
\label{lmd}
\lambda_{0}=\frac{n-2}{2n},
\qquad
\mu_{0}=\frac{n+2}{2n}
\qquad
\text{and}
\qquad
\delta_{0}=\mu_{0}-\lambda_{0}=\frac{2}{n}.
\end{gather}
Thus, the data of a~conformal manifold $(M,[{\rm g}])$ is enough to def\/ine
$\Delta_{Y}\in\mathcal{D}_{\lambda_0,\mu_0}(M)$.
We write it below as $\Delta_{Y}^M({\rm g})$ and we refer to it as the Yamabe or conformal Laplacian.
One easily gets
{\samepage \begin{Proposition}
\label{prop:Yamabe}
The conformal Laplacian is a~natural conformally invariant operator, i.e.
\begin{itemize}\itemsep=0pt
\item it satisfies the naturality condition:
\begin{gather}
\label{nat0}
\Delta_{Y}^N(\Phi^*{\rm g})=\Phi^*\left(\Delta_{Y}^M({\rm g})\right),
\end{gather}
for all diffeomorphisms $\Phi:
N\rightarrow M$ and for all pseudo-Rie\-man\-nian metric ${\rm g}$ on $M$,
\item it is conformally invariant, $\Delta_{Y}^M(e^{2\Upsilon}{\rm g})=\Delta_{Y}^M({\rm g})$ for all
$\Upsilon\in\mathcal{C}^{\infty}(M)$.
\end{itemize}
\end{Proposition}}

More generally, a~natural operator over pseudo-Rie\-man\-nian manifolds is an operator that acts between
natural bundles, is def\/ined over any pseudo-Rie\-man\-nian manifold $(M,{\rm g})$ and satisf\/ies an analogue
of the naturality condition~\eqref{nat0}.
It is said to be conformally invariant if it depends only on the conformal class of ${\rm g}$.
For a~general study of natural operators in the pseudo-Rie\-man\-nian setting, see the book~\cite{KMS93}.

From Proposition~\ref{prop:Yamabe}, we deduce that the conformal Laplacian $\Delta_{Y}$ is invariant under
the action of conformal dif\/feomorphisms, which reads inf\/initesimally as
\begin{gather}
\label{InvConfDel}
L^{\mu_{0}}_X\circ\Delta_{Y}=\Delta_{Y}\circ L^{\lambda_{0}}_X,
\end{gather}
for all conformal Killing vector f\/ields $X$.
Here, as introduced in~\eqref{LieDerivative}, $L^{\lambda_{0}}$ and $L^{\mu_{0}}$ denote the Lie
derivatives of $\lambda_{0}$- and $\mu_{0}$-densities.
If the manifold $(M,[{\rm g}])$ is locally conformally f\/lat, then, up to multiplication by a~scalar,
$\Delta_{Y}$ is the unique second order operator acting on densities which is invariant under the
action~\eqref{InvConfDel} of conformal Killing vector f\/ields.

\subsection{Natural and conformally invariant quantization}

Recall f\/irst the def\/inition of a~quantization on a~smooth manifold $M$.
\begin{Definition}
Let $\lambda,\mu\in\mathbb{R}$ and $\delta=\mu-\lambda$.
A quantization on $M$ is a~linear bijection $\mathcal{Q}_{\lambda,\mu}^{M}$ from the space of symbols
$\mathcal{S}_{\delta}(M)$ to the space of dif\/ferential operators $\mathcal{D}_{\lambda,\mu}(M)$ such that
\begin{gather*}
\sigma_k\big(\mathcal{Q}_{\lambda,\mu}^{M}(S)\big)=S,
\qquad
\forall\, S\in\mathcal{S}_{\delta}^{k}(M),
\quad
\forall\, k\in\mathbb{N}.
\end{gather*}
\end{Definition}

On locally conformally f\/lat manifolds $(M,[{\rm g}])$, for generic weights $\lambda$, $\mu$, there exists
a~unique conformally equivariant quantization~\cite{DLO99}, i.e.\ a unique quantization which intertwines
the actions of the conformal Killing vector f\/ields on $\mathcal{S}_\delta(M)$ and on
$\mathcal{D}_{\lambda,\mu}(M)$.
In the following, we need an extension of the conformally equivariant quantization to arbitrary conformal
manifolds.
This is provided by the notion of {\it natural and conformally invariant quantization}.
The def\/inition and the conjecture of the existence of such a~quantization were given for the f\/irst time
in~\cite{Lec01}.
\begin{Definition}
A natural and conformally invariant quantization is the data for every pseudo-Rie\-man\-nian manifold $(M,{\rm
g})$ of a~quantization $\mathcal{Q}_{\lambda,\mu}^M({\rm g})$, which satisf\/ies
\begin{itemize}\itemsep=0pt
\item the naturality condition:
\begin{gather}
\label{nat}
\mathcal{Q}_{\lambda,\mu}^{N}(\Phi^{*}{\rm g})(\Phi^{*}S)=\Phi^{*}(\mathcal{Q}_{\lambda,\mu}^{M}({\rm g})(S)),
\qquad
\forall\, S\in\mathcal{S}_{\delta}(M),
\end{gather}
for all dif\/feomorphisms $\Phi:
N\rightarrow M$ and for all pseudo-Rie\-man\-nian metric ${\rm g}$ on $M$.
\item the conformal invariance: $\mathcal{Q}_{\lambda,\mu}^M(e^{2\Upsilon}{\rm
g})=\mathcal{Q}_{\lambda,\mu}^M({\rm g})$ for all $\Upsilon\in\mathcal{C}^{\infty}(M)$.
\end{itemize}
\end{Definition}
In the following we refer to a~quantization map by $\mathcal{Q}_{\lambda,\mu}$, the dependence in the
chosen pseudo-Rie\-man\-nian manifold $(M,{\rm g})$ being understood.
Accordingly, we drop the reference to~$M$ in the spaces of densities $\mathcal{F}_\lambda$, symbols
$\mathcal{S}_\delta$ and dif\/ferential operators $\mathcal{D}_{\lambda,\mu}$.

The concept of natural and conformally invariant quantization is an extension to quantizations of the more
usual one of natural conformally invariant operator, introduced in the previous section.
Restricting to conformally f\/lat manifolds $(M,[{\rm g}])$ and to $\Phi\in\mathrm{Dif\/f}(M)$ preserving~$[{\rm g}]$,
the naturality condition~\eqref{nat} reads as conformal equivariance of the quantization map
$\mathcal{Q}_{\lambda,\mu}$.
Thus, the problem of the natural and conformally invariant quantization on an arbitrary ma\-ni\-fold
generalizes the problem of the conformally equivariant quantization on conformally f\/lat manifolds.

Remark that the bundles $S^{k}TM$ are natural bundles over $(M,[{\rm g}])$.
Hence, one can consider natural and conformally invariant quantization restricted to the subspaces of
symbols $\mathcal{S}_\delta^{k}$ or $\mathcal{S}_\delta^{\leq k}=\bigoplus_{j\leq k}\mathcal{S}^j_\delta$.
In a~f\/irst step, the proofs of the existence of a~natural and conformally invariant quantization at the
second and the third orders were given respectively in~\cite{DOv01} and~\cite{Lou03}, together with
explicit formulas.
We provide the one at order~$2$, which we will need later on.
\begin{Theorem}
[\cite{DOv01}]
\label{thm:ExplicitQ}
Let $\delta\notin\left\{\frac{2}{n},\frac{n+2}{2n},1,\frac{n+1}{n},\frac{n+2}{n}\right\}$.
A natural and conformally invariant quantization $\mathcal{Q}_{\lambda,\mu}:
\mathcal{S}^{\leq 2}_\delta\rightarrow\mathcal{D}_{\lambda,\mu}^{2}$ is provided, on a~pseudo-Rie\-man\-nian manifold $(M,{\rm
g})$ of dimension~$n$, by the formulas
\begin{gather}
\nonumber
\mathcal{Q}_{\lambda,\mu}(f) = f,
\\
\nonumber
\mathcal{Q}_{\lambda,\mu}(X) = X^a\nabla_a+\frac{\lambda}{1-\delta}(\nabla_aX^a),
\\
\mathcal{Q}_{\lambda,\mu}(S) = S^{ab}\nabla_a\nabla_b+\beta_1(\nabla_aS^{ab}
)\nabla_b+\beta_2{\rm g}^{ab}(\nabla_a\Tr S)\nabla_b
\nonumber
\\
\phantom{\mathcal{Q}_{\lambda,\mu}(S)=}
+\beta_3\big(\nabla_a\nabla_bS^{ab}\big)+\beta_4{\rm g}^{ab}\nabla_a\nabla_b(\Tr S),
+\beta_5\Ric_{ab}S^{ab}+\beta_6\Sc(\Tr S),
\label{Formula:Q}
\end{gather}
where $f$, $X$, $S$ are symbols of degrees $0$, $1$, $2$ respectively and $\Tr S={\rm g}_{ab}S^{ab}$.
Moreover the coefficients $\beta_i$ entering the last formula are given by
\begin{gather}
\beta_1=\frac{2(n\lambda+1)}{2+n(1-\delta)},\nonumber
\\
\beta_2=\frac{n(\lambda+\mu-1)}{(2+n(1-\delta))(2-n\delta)},\nonumber
\\
\beta_3=\frac{n\lambda(n\lambda+1)}{(1+n(1-\delta))(2+n(1-\delta))},\nonumber
\\
\beta_4=\frac{n\lambda(n^2\mu(2-\lambda-\mu)+2(n\lambda+1)^2-n(n+1))}
{(1+n(1-\delta))(2+n(1-\delta))(2+n(1-2\delta))(2-n\delta)},\nonumber
\\
\beta_5=\frac{n^2\lambda(\mu-1)}{(n-2)(1+n(1-\delta))},\nonumber
\\
\beta_6=\frac{n^2\lambda(\mu-1)(n\delta-2)}{(n-1)(n-2)(1+n(1-\delta))(2+n(1-2\delta))}.\label{beta}
\end{gather}
\end{Theorem}

In a~second step, the proof of the existence of such a~quantization, at an arbitrary order and for generic
values of $\lambda$, $\mu$, was given in~\cite{CSi09,MRa10b,Sil09} in dif\/ferent ways.
We provide a~slightly ref\/ined statement in the next section.

\subsection{Adjoint operation and quantization}

For all weights $\lambda\in\mathbb{R}$, there exists a non-degenerate symmetric bilinear pairing
\begin{alignat*}{4}
& \mathcal{F}^c_\lambda\times\mathcal{F}^c_{1-\lambda}
&&
\rightarrow \ &&
 \mathbb{R}, &
\\
& (\varphi,\psi)
&&
\mapsto \
&&
\int_M\varphi\psi,&
\end{alignat*}
where $\mathcal{F}^c_\lambda$ is the space of compactly supported $\lambda$-densities.
On a~manifold $M$, this pairing is $\mathrm{Dif\/f}(M)$-invariant since $1$-density is the right object for
integration.
In consequence, we can def\/ine an adjoint operation
$*:
\mathcal{D}_{\lambda,\mu}
\rightarrow
\mathcal{D}_{1-\mu,1-\lambda}$ by
\begin{gather*}
\big(\varphi,D^*\psi\big)=(D\varphi,\psi),
\end{gather*}
for all $\varphi\in\mathcal{F}^c_\lambda$ and $\psi\in\mathcal{F}^c_{1-\mu}$.
We introduce the following subset of $\mathbb{R}^2$,
\begin{gather*}
I=\left\{(\lambda,\mu)\in\mathbb{R}^2\,\big|\,
\mu-\lambda\notin\frac{1}{2n}(\mathbb{N}\setminus\{0\})\right\}
\cup\{(\lambda_{0},\mu_{0})\},
\end{gather*}
where $\lambda_{0}=\frac{n-2}{2n}$ and $\mu_{0}=\frac{n+2}{2n}$ are the weights of the conformal Laplacian
(see~\eqref{lmd}).
The set $I$ is stable under the involutive map $(\lambda,\mu)\mapsto (1-\mu,1-\lambda)$.
Note that
\begin{gather*}
\sigma_2(\Delta_{Y})=|\mathrm{Vol}_{\rm g}|^{\delta_0}H,
\end{gather*}
where $\delta_0=\mu_{0}-\lambda_{0}$ and $H={\rm g}^{ij}p_ip_j$ in canonical coordinates.
The proof of existence of a~natural and conformally invariant quantization $\mathcal{Q}_{\lambda,\mu}$
in~\cite{Sil09} leads easily to the following statement.
\begin{Theorem}
\label{Thm:NCIQ}
There exists a~family $(\mathcal{Q}_{\lambda,\mu})_{(\lambda,\mu)\in I}$ of natural and conformally
invariant quantizations that satisfies:
\begin{itemize}
\itemsep=0pt \item the reality condition:
\begin{gather}
\label{real}
\mathcal{Q}_{\lambda,\mu}(S)^*=(-1)^k\mathcal{Q}_{1-\mu,1-\lambda}(S),
\qquad
\forall\, S\in\mathcal{S}_{\delta}^{k},
\qquad
\forall\,(\lambda,\mu)\in I,
\end{gather}
\item the factorization property:
\begin{gather}
\mathcal{Q}_{\lambda_{0},\lambda_{0}}\big(|\mathrm{Vol}_{\rm g}|^{\delta_0}HS\big) = \mathcal{Q}
_{\mu_{0},\lambda_{0}}(S)\circ\Delta_{Y},
\qquad
\forall\, S\in\mathcal{S}_{-\delta_0}^{k},
\nonumber
\\
\mathcal{Q}_{\mu_{0},\mu_{0}}\big(|\mathrm{Vol}_{\rm g}|^{\delta_0}HS\big) = \Delta_{Y}
\circ\mathcal{Q}_{\mu_{0},\lambda_{0}}(S),
\qquad
\forall\, S\in\mathcal{S}_{-\delta_0}^{k},
\label{factorization}
\end{gather}
\item the restriction of $\mathcal{Q}_{\lambda,\mu}$ to $\mathcal{S}_\delta^{\leq 2}$ is given by the
formulas in~\eqref{Formula:Q} if $(\lambda,\mu)\in I\setminus\{(\lambda_{0},\mu_{0})\}$.
\end{itemize}
\end{Theorem}
\begin{proof}
We prove the theorem in four steps.

In~\cite[Theorem 4.4]{Sil09}, one of us determines that for $(\lambda,\mu)\in I$ there exists a~natural and
conformally invariant quantization map $\mathcal{Q}_{\lambda,\mu}'''$.

From the above family of quantizations $(\mathcal{Q}_{\lambda,\mu}''')_{(\lambda,\mu)\in I}$, we def\/ine
$(\mathcal{Q}_{\lambda,\mu}'')_{(\lambda,\mu)\in I}$ by
\begin{gather*}
\mathcal{Q}_{\lambda,\mu}'':
\
S
\
\mapsto
\
\frac{1}{2}\left(\mathcal{Q}_{\lambda,\mu}'''(S)+(-1)^k\mathcal{Q}'''_{1-\mu,1-\lambda}(S)^*\right),
\qquad
\forall\, S\in\mathcal{S}_\delta^k,
\quad
\forall\,(\lambda,\mu)\in I.
\end{gather*}
The maps $\mathcal{Q}_{\lambda,\mu}''$ are again natural and conformally invariant quantizations.
Indeed, the adjoint operation $*$ is natural, does not depend of the choice of metric on $M$ and satisf\/ies
\begin{gather*}
\sigma_k\big(D^*\big)=(-1)^k\sigma_k(D),
\end{gather*}
for all dif\/ferential operators $D$ of order $k$.
The newly def\/ined quantization maps clearly satisfy the property~\eqref{real} since $*$ is an involution.

For $(\lambda,\mu)\in I\setminus \{(\lambda_{0},\lambda_{0}),(\mu_{0},\mu_{0})\}$, we def\/ine
$\mathcal{Q}_{\lambda,\mu}':=\mathcal{Q}_{\lambda,\mu}''$.
On the space of traceless symbols we set
$\mathcal{Q}'_{\lambda_{0},\lambda_{0}}:=\mathcal{Q}''_{\lambda_{0},\lambda_{0}}$ and
$\mathcal{Q}'_{\mu_{0},\mu_{0}}:=\mathcal{Q}''_{\mu_{0},\mu_{0}}$.
We extend both maps to the whole symbol space by the formulas in~\eqref{factorization}.
They are clearly still natural and conformally invariant and satisfy the reality condition~\eqref{real}.

For all $(\lambda,\mu)\in I\setminus\{(\lambda_{0},\mu_{0})\}$, we denote by $\mathcal{Q}_{\lambda,\mu}$
the natural and conformally invariant quantizations restricted to $\mathcal{S}_\delta^{\leq 2}$, given by
the formulas~\eqref{Formula:Q}.
A direct computation shows that they satisfy the reality condition~\eqref{real} and the factorization
property~\eqref{factorization}.
For $(\lambda,\mu)=(\lambda_{0},\mu_{0})$, we set
$\mathcal{Q}_{\lambda_{0},\mu_{0}}:=\mathcal{Q}'_{\lambda_{0},\mu_{0}}$ on $\mathcal{S}_{\delta_0}^{\leq
2}$.
We extend then the quantizations $\mathcal{Q}_{\lambda,\mu}$ (where \mbox{$(\lambda,\mu)\in I$}) to the whole
symbol space by setting $\mathcal{Q}_{\lambda,\mu}:=\mathcal{Q}_{\lambda,\mu}'$ on
$\mathcal{S}_\delta^{\geq 3}$, for all $(\lambda,\mu)\in I$.
\end{proof}

In the following, the quantization maps $\mathcal{Q}_{\lambda,\mu}$ that we will use are always taken from
a~family $(\mathcal{Q}_{\lambda,\mu})_{(\lambda,\mu)\in I}$ provided by Theorem~\ref{Thm:NCIQ}.
In fact, we will need only four of them, namely: $\mathcal{Q}_{\lambda_{0},\lambda_{0}}$,
$\mathcal{Q}_{\mu_{0},\mu_{0}}$, $\mathcal{Q}_{\lambda_{0},\mu_{0}}$, $\mathcal{Q}_{\mu_{0},\lambda_{0}}$.
With such a~convention, it is worth noticing that the conformal Laplacian can be obtained as
\begin{gather*}
\Delta_{Y}=\mathcal{Q}_{\lambda_{0},\mu_{0}}\big(|\mathrm{Vol}_{\rm g}|^{\delta_0}H\big).
\end{gather*}
The conformal invariance of the symbol $|\mathrm{Vol}_{\rm g}|^{\delta_0}H$ translates into the conformal
invariance of~$\Delta_{Y}$.

\section{On particular conformally invariant operators}\label{Section3}

First, we introduce notation for classical objects of the pseudo-Rie\-man\-nian and conformal geo\-metries and
recall basic facts about natural and conformally invariant operators.
Then, we classify the natural conformally invariant operators between particular subspaces of symbols.

\subsection{More on pseudo-Rie\-man\-nian and conformal geometry}
We complete here Section~\ref{Par:notation0}, and use freely the notation introduced there.

First, we work over a~pseudo-Rie\-man\-nian manifold.
The Riemann tensor admits the following decomposition
\begin{gather}
\label{Decompo-R}
\mathrm{R}_{ab}{}^c{}_d=\mathrm{C}_{ab}{}^c{}_d+2\delta^c_{[a}\mathrm{P}_{b]d}^{}+2{\rm g}_{d[b}\mathrm{P}_{a]}{}^c,
\end{gather}
where $\mathrm{C}_{ab}{}^c{}_d$ is the totally trace-free Weyl curvature,
$\mathrm{P}_{ab}=\frac{1}{n-2}\left(\Ric_{ab}-\frac{1}{2(n-1)}\Sc{\rm g}_{ab}\right)$ is the
Schouten tensor, $\delta_a^b$ is the Kronecker delta and square brackets denote antisymmetrization of
enclosed indices.
The Weyl tensor $\mathrm{C}_{abcd}$ is zero for the dimension $n=3$.
Note also that $\mathrm{C}_{abcd}$ obeys the same symmetries of indices as $\mathrm{R}_{abcd}$ does.
Further curvature quantities we shall need are
\begin{gather*}
\mathrm{J}={\rm g}^{ab}\mathrm{P}_{ab}
\qquad
\text{and}
\qquad
\mathrm{A}_{abc}=2\nabla_{[b}\mathrm{P}_{c]a},
\end{gather*}
where $\mathrm{A}_{abc}$ is the Cotton--York tensor and $\mathrm{J}$ is related to the scalar curvature via
$\mathrm{J}=\frac{1}{2(n-1)}\Sc$.
Bianchi identities have the from $\mathrm{R}_{[abc]d}=0$ and $\nabla_{[a}\mathrm{R}_{bc]de}=0$ and lead to
\begin{gather*}
(n-3)\mathrm{A}_{abc}=\nabla_r\mathrm{C}_{bc}{}^r{}_a
\qquad
\text{and}
\qquad
\nabla_b\mathrm{P}^{b}{}_a=\nabla_a\mathrm{J}.
\end{gather*}

Second, we consider a~conformal manifold $(M,[{\rm g}])$.
The Weyl tensor $\mathrm{C}_{ab}{}^c{}_d$ is a~conformal invariant, i.e.\ it does not depend on the choice
of the representative metric from $[{\rm g}]$.
The same is true for $\mathrm{A}_{abc}$ in the dimension 3.
Further, a~choice of metric provides a~canonical trivialization of the bundle of $\lambda$-densities
$F_{\lambda}$ via the global section $|\mathrm{Vol}_{\rm g}|^\lambda$ (see Section~\ref{basic}).
According to the transformation rule $|\mathrm{Vol}_{\hat{{\rm g}}}|^\lambda=e^{n\lambda\Upsilon}|\mathrm{Vol}_{{\rm g}}|$
if $\hat{{\rm g}}=e^{2\Upsilon}{\rm g}$,
we have the conformally invariant object ${\bold g}_{ab}={\rm g}_{ab}\otimes|\mathrm{Vol}_{{\rm g}}|^{-\frac{2}{n}}$,
termed {\em conformal metric}, with the inverse ${\bold g}^{ab}$ in $\Gamma(S^2TM) \otimes
\mathcal{F}_{2/n}$, see e.g.~\cite{BEG94} for details.
(Note that the space of densities $\mathcal{E}[w]$ in~\cite{BEG94} corresponds to $\mathcal{F}_{-w/n}$ in
our notation.) The conformal metric gives a~conformally invariant identif\/ication $TM \cong T^*M \otimes
F_{-2/n}$.
In other words, we can raise and lower indices, with expense of the additional density, in a~conformally
invariant way.
For example, we get $\mathrm{C}_{abcd} \in \Gamma(S^2(\Lambda^2T^*M)) \otimes \mathcal{F}_{-2/n}$.
Note also that ${\bold g}_{ab}$ and ${\bold g}^{ab}$ are parallel for any choice of a~Levi-Civita
connection from the conformal class.

\subsection{Description of natural conformally invariant operators}

Now we recall basic facts about natural
and conformally invariant operators.
Every natural operator on the Riemannian structure $(M,{\rm g})$ between natural bundles $V_1$ and $V_2$ is
a~linear combination of terms of the form
\begin{gather}
\label{natural}
\underbrace{{\rm g}^{-1}\cdots{\rm g}^{-1}}_{r_1}\underbrace{{\rm g}\cdots{\rm g}}_{r_2}
\underbrace{(\nabla^{(i_1)}\mathrm{R})\cdots(\nabla^{(i_s)}\mathrm{R})}_s\underbrace{\nabla\cdots\nabla}_t f
\end{gather}
to which one applies a~${\rm GL}(n)$-invariant operation
\begin{gather}
\label{naturalbis}
\Gamma\big(\bigotimes{}^{2r_1+s}TM\otimes\bigotimes{}^{2r_2+3s+i+t}T^*M\otimes V_1\big)
\
\longrightarrow
\
\Gamma(V_2).
\end{gather}
Here $f \in \Gamma(V_1)$, ${\rm g}^{-1}$ stands for the inverse of the metric ${\rm g}$, $\nabla^{(i_j)}$
denotes the $i_j$th iterated covariant derivative where $i_j \geq 0$, abstract indices are omitted, $i =
i_1 + \cdots +i_s$, and $\nabla$ and $\mathrm{R}$ correspond to the choice of the metric ${\rm g}$.
The existence of a~${\rm GL}(n)$-invariant operation~\eqref{naturalbis} gives in general constraints on the
possible values of $r_1$, $r_2$, $s$, $t$, $i$.
See~\cite{KMS93} for details.

A natural operator on $(M,{\rm g})$ is conformally invariant if it does not depend on the choice of metric
in the conformal class.
Then, it def\/ines a~natural operator on the conformal structure $(M,[{\rm g}])$.
It is convenient to use the conformal metric ${\bold g}$ instead of ${\rm g}$ and the inverse ${\bold
g}^{-1}$ instead of ${\rm g}^{-1}$ in~\eqref{natural} since they are conformally invariant, namely
\begin{gather}
\label{naturalC}
\underbrace{{\bold g}^{-1}\cdots{\bold g}^{-1}}_{r_1}\underbrace{{\bold g}\cdots{\bold g}}_{r_2}
\underbrace{(\nabla^{(i_1)}\mathrm{R})\cdots(\nabla^{(i_s)}\mathrm{R})}_s\underbrace{\nabla\cdots\nabla}_t f
\nonumber
\\
\qquad{}
\in\Gamma\left(\bigotimes{}^{2r_1+s}TM\otimes\bigotimes{}^{2r_2+3s+i+t}
T^*M\otimes V_1\otimes F_{\frac2n(r_1-r_2)}\right)
\end{gather}
for $f \in \Gamma(V_1)$.
It is generally a~dif\/f\/icult problem to determine which linear combinations of terms as
in~\eqref{naturalC}, together with suitable projections as in~\eqref{naturalbis}, give rise to
a~conformally invariant operator.
We shall need details only in specif\/ic cases.

\subsection{Conformally invariant operators on the symbol space} This section concerns existence and
uniqueness of natural and conformally invariant operators of certain type.
The f\/irst one is well-known and can be obtained as an easy consequence of~\cite{Feg76}, or deduced from
the general work~\cite{CSS01} on curved BGG-sequences.
We present a~detailed proof to demonstrate the technique which is used (in much more complicated setting)
later in the proof of Proposition~\ref{Prop:Obs}.

Recall f\/irst that $\Gamma(S^{k,0} TM)$ is the space of trace-less symmetric $k$-tensors.
In terms of the abstract index notation, a~section $f$ of $\bigotimes^k TM$ is denoted by $f^{a_1\dots a_k}$.
In the following, we write $f^{[a_1\dots a_k]}$, $f^{(a_1\dots a_k)}$ and $f^{(a_1\dots a_k)_0}$ for the
projections of $f$ to $\Gamma(\Lambda^k TM)$, $\Gamma(S^k TM)$ and $\Gamma(S^{k,0} TM)$, respectively.
Similar notation will be used for covariant indices.
\begin{Proposition}
\label{Prop:G}
Up to multiplication by a~scalar, there exists a~unique natural conformally invariant operator
$\mathcal{S}^k_0\rightarrow\mathcal{S}^{k+1}_{2/n}$.
It is given by the conformal Killing operator $\mathbf{G}$, such that for all $f\in\mathcal{S}^k_0$,
\begin{gather}
\label{ConfInvG}
(\mathbf{G}(f))^{a_0\dots a_k}=\nabla^{(a_0}f^{a_1\dots a_k)_0}.
\end{gather}
\end{Proposition}
\begin{proof}
Identifying $\mathcal{S}^k_0$ and $\mathcal{S}^{k+1}_{2/n}$ with corresponding spaces of sections of
symmetric tensors, we consider natural and conformally invariant operators $\Gamma(S^kTM) \to
\Gamma(S^{k+1}TM \otimes F_{2/n})$.
By naturality, such operators are linear combinations of terms in~\eqref{naturalC} with $V_1=S^kTM$,
composed with ${\rm GL}_n$-invariant maps
\begin{gather*}
\bigotimes{}^{2r_1+s}TM\otimes\bigotimes{}^{2r_2+3s+i+t}T^*M\otimes S^kTM\otimes F_{\frac2n(r_1-r_2)}
\
\longrightarrow
\
S^{k+1}TM\otimes F_{2/n}.
\end{gather*}
Explicitly, those maps may consist of: contracting covariant and contravariant indices, projecting the
covariant and contravariant tensors on tensors of prescribed symmetry type (given by a~Young diagram) and
tensorizing with the density $|\mathrm{Vol}_{\rm g}|^\delta$ for arbitrary $\delta\in\mathbb{R}$.
The conformal invariance does not allow for the last operation, hence $r_1 - r_2 =1$.
The dif\/ference between the number of covariant and contravariant indices is a~constant therefore
$(2r_1+s+k) - (2r_2+3s+i+t)=k+1$, i.e.\ $2s+i+t=1$.
This means $s=i=0$ and $t=1$.
The sought operators are then f\/irst order (gradient) natural operators and using moreover the conformal
invariance, the statement follows from the classif\/ication in~\cite{Feg76}.
\end{proof}

The next proposition is a~crucial technical tool in the following.
\begin{Proposition}
\label{Prop:Obs}
Every natural conformally invariant operator $\mathcal{S}^{k,0}_0 \rightarrow \mathcal{S}^{k-1}_{2/n}$ has
its target space in $\mathcal{S}^{k-1,0}_{2/n} \subseteq \mathcal{S}^{k-1}_{2/n}$.
The space of natural conformally invariant operators $\mathcal{S}^{k,0}_0 \to \mathcal{S}^{k-1,0}_{2/n}$ on
a~pseudo-Rie\-man\-nian manifold $(M,{\rm g})$ is at most two-dimensional and depends on $k \in \mathbb{N}$ as
follows.
\begin{enumerate}\itemsep=0pt
\item[$(i)$] This space is trivial for $k=1$.

\item[$(ii)$] If $k=2$ or $n=3$,
this space is one-dimensional and generated by the operator $\mathbf{F}$ such that, for all $f\in\mathcal{S}^{k,0}_{0}$,
\begin{gather*}
(\mathbf{F}(f))^{a_1\ldots a_{k-1}}=\mathrm{C}^r{}_{st}{}^{(a_1}\nabla_r f^{a_2\ldots a_{k-1})_0st}
-(k+1)\mathrm{A}_{st}{}^{(a_1}f^{a_2\ldots a_{k-1})_0st}.
\end{gather*}

\item[$(iii)$]  If $k=3$ and $n>3$,
this space is two-dimensional and generated by two operators, $\mathbf{F}_{1}$~and~$\mathbf{F}_{2}$, such that for all $f\in\mathcal{S}^{k,0}_0$,
\begin{gather*}
(\mathbf{F}_{1}(f))^{a_1\ldots a_{k-1}} = (\mathbf{F}(f))^{a_1\ldots a_{k-1}}+\frac{k-2}
{n+2k-2}\mathrm{C}^{(a_1}{}_r{}^{a_2}{}_s\nabla_t f^{a_3\ldots a_{k-1})_0rst},
\\
(\mathbf{F}_{2}(f))^{a_1\ldots a_{k-1}} = 4\mathrm{C}^{(a_1}{}_r{}^{a_2}{}_s\nabla_t f^{a_3\ldots a_{k-1})_0rst}
+(n+2k-2)(\nabla_r\mathrm{C}_s{}^{(a_1}{}_t{}^{a_2})f^{a_3\ldots a_{k-1})_0rst}
\\
\phantom{(\mathbf{F}_{2}(f))^{a_1\ldots a_{k-1}} =}{}
+2(n+2k-2)\mathrm{A}_{rs}{}^{(a_1}f^{a_2\ldots a_{k-1})_0rs}.
\end{gather*}
\end{enumerate}
\end{Proposition}

\begin{Remark}
Let $(x^i,p_i)$ be a~canonical coordinate system on $T^*M$.
We can then write the operators $\mathbf{G}$ and $\mathbf{F}$ as follows on $\mathcal{S}^k_0$
\begin{gather}
\label{ConfInv-p}
\mathbf{G}=\Pi_0\circ\big({\bold g}^{ij}p_i\nabla_j\big)
\qquad
\text{and}
\qquad
\mathbf{F}=\Pi_0\circ{\bold g}^{im}p_i\partial_{p_j}\partial_{p_l}\big(\mathrm{C}^k{}_{jlm}
\nabla_k-(k+1)\mathrm{A}_{jlm}\big),
\end{gather}
where $\Pi_0:
\mathcal{S}^{k-1}_0\rightarrow\mathcal{S}^{k-1,0}_{2/n}$ is the canonical projection on
trace-less symbols.
Actually, we will see in the sequel that the conformal Killing operator $\mathbf{G}$ can be used to
def\/ine the conformal Killing tensors whereas the operator $\mathbf{F}$ occurs in the computation of the
obstruction to the existence of conformal symmetries of $\Delta_{Y}$.
\end{Remark}

Let us note that the proof of Proposition~\ref{Prop:Obs} is long, technical and interesting rather for
experts in conformal geometry.
The reader interested mainly in results about symmetries can continue the reading in
Section~\ref{sym2nd} (details from the proof will not be needed there).
\begin{proof}
We study natural and conformally invariant operators $\Gamma(S^{k,0}TM) \!\to\! \Gamma(S^{k-1}TM {\otimes}
F_{2/n})$.
In the f\/irst part of the proof we consider the naturality and in the second part the conformal invariance.

I.\ {\bf Naturality.} We start in a~similar way as in the proof of Proposition~\ref{Prop:G}.
By naturality, the considered operators are linear combinations of terms in~\eqref{naturalC} composed with
$\mathrm{GL}_n$-invariant maps
\begin{gather*}
\bigotimes{}^{2r_1+s}TM\otimes\bigotimes{}^{2r_2+3s+i+t}T^*M\otimes S^kTM\otimes F_{\frac2n(r_1-r_2)}
\
\longrightarrow
\
S^{k-1}TM\otimes F_{2/n}.
\end{gather*}
The conformal invariance of the discussed operators leads to $r_1 - r_2 =1$ and the ${\rm GL}_n$-invariance of
the maps above imposes $(2r_1+s+k) - (2r_2+3s+i+t)=k-1$, i.e.\ $2s+i+t=3$.
This means either $s=i=0$, $t=3$ or $s=i=1$, $t=0$ or $s=t=1$, $i=0$.
Hence, omitting abstract indices, the natural operators $\Gamma(S^{k,0}TM) \to \Gamma(S^{k-1}TM \otimes
F_{2/n})$ are a~linear combination of terms
\begin{gather}
\label{terms}
\underbrace{{\bold g}^{-1}\cdots{\bold g}^{-1}}_{r+1}\underbrace{{\bold g}\cdots{\bold g}}_{r}\nabla\nabla\nabla f,
\qquad
\underbrace{{\bold g}^{-1}\cdots{\bold g}^{-1}}_{r+1}\underbrace{{\bold g}\cdots{\bold g}}_{r}\mathrm{R}\nabla f,
\qquad
\underbrace{{\bold g}^{-1}\cdots{\bold g}^{-1}}_{r+1}\underbrace{{\bold g}\cdots{\bold g}}_{r}(\nabla\mathrm{R})f,
\end{gather}
where $r \geq 0$ and $f \in \Gamma(S^{k,0}TM)$, each of which is followed by a~${\rm GL}(n)$-invariant
projection to $\Gamma(S^{k-1}TM \otimes F_{2/n})$.
Irreducible components of the target bundle $S^{k-1}TM \otimes F_{2/n}$ are
\begin{gather*}
S^{k-1,0}TM\otimes F_{2/n},
\qquad
S^{k-3,0}TM,
\qquad
S^{k-5,0}TM\otimes F_{-2/n}, \qquad \ldots,
\end{gather*}
but since $f$ is trace-free, one easily verif\/ies from~\eqref{terms} that only possible
target bundles are $S^{k-1,0}TM \otimes F_{2/n}$ and $S^{k-3,0}TM$.
In other words, in the expressions~\eqref{terms}, one can restrict to $r=0$.

It remains to describe possible ${\rm GL}(n)$-invariant projections of the terms in~\eqref{terms} in details.
Using the decomposition~\eqref{Decompo-R} of $\mathrm{R}$ into Weyl and Schouten tensors, they split into
f\/ive terms: ${\bold g}^{-1}\nabla\nabla\nabla f$, ${\bold g}^{-1}\mathrm{C}\nabla f$, ${\bold
g}^{-1}(\nabla\mathrm{C}) f$, ${\bold g}^{-1}\mathrm{P}\nabla f$ and ${\bold g}^{-1}(\nabla\mathrm{P}) f$.

We shall start with natural operators $\Gamma(S^{k,0}TM) \to \Gamma(S^{k-1,0}TM \otimes F_{2/n})$.
In this situation, at least one of the two indices above ${\bold g}^{-1}$ in the expression of the operator
has to be contracted with a~covariant index.
For an operator of type ${\bold g}^{-1}\nabla\nabla\nabla f$, the two resulting operators are (up to the
order of covariant derivatives) respectively
\begin{gather}
\label{nabla3}
\nabla^r\nabla_r\nabla_s f^{a_1\ldots a_{k-1}s},
\qquad
\nabla^{(a_1}\nabla_{s}\nabla_t f^{a_2\ldots a_{k-1})_0st}.
\end{gather}
Since the change of the order of covariant derivatives gives rise to curvature operators of the form
${\bold g}^{-1}\mathrm{R}\nabla f$ and ${\bold g}^{-1}(\nabla\mathrm{R}) f$, the previous display is
suf\/f\/icient for operators of type ${\bold g}^{-1}\nabla\nabla\nabla f$.
Using that $\mathrm{C}$ is completely trace-free and $(n-3)\mathrm{A}_{abc}=\nabla_r
\mathrm{C}_{bc}{}^r{}_a$, the dif\/ferent possibilities of contraction of indices for the expressions
${\bold g}^{-1}\mathrm{C}\nabla f$ and ${\bold g}^{-1}(\nabla\mathrm{C})f$ lead to the operators
\begin{gather}
\mathrm{C}^{r}{}_{st}{}^{(a_1}\nabla_r f^{a_2\ldots a_{k-1})_0st},
\qquad
\mathrm{C}^{(a_1}{}_r{}^{a_2}{}_s\nabla_t f^{a_3\ldots a_{k-1})_0rst},
\nonumber
\\
\big(\nabla_r\mathrm{C}_s{}^{(a_1}{}_t{}^{a_2})f^{a_3\ldots a_{k-1}\big)_0rst},
\qquad
\mathrm{A}_{rs}{}^{(a_1}f^{a_2\ldots a_{k-1})_0rs}.
\label{nablaC}
\end{gather}
Thanks to the decomposition of $\mathrm{P}$ into irreducible components and to the equality
$\nabla_{a}\mathrm{P}^{a}{}_{b}=\nabla_{b}J$, we see that the dif\/ferent conf\/igurations of indices in
the expressions ${\bold g}^{-1}\mathrm{P}\nabla f$ and ${\bold g}^{-1}(\nabla\mathrm{P})f$ give rise to the
operators
\begin{gather}
\mathrm{P}_{(rs)_0}\nabla^{r}f^{sa_1\ldots a_{k-1}},
\qquad
\mathrm{P}_{(rt)_0}{\bold g}^{t(a_1}\nabla_s f^{a_2\ldots a_{k-1})_0rs},
\qquad
\mathrm{P}_{rs}\nabla^{(a_1}f^{a_2\ldots a_{k-1})_0rs},
\nonumber
\\
\mathrm{J}\nabla_r f^{a_1\ldots a_{k-1}r},
\qquad
(\nabla_{(r}\mathrm{P}_{st)_0}){\bold g}^{t(a_1}f^{a_2\ldots a_{k-1})_0rs},
\qquad
(\nabla_r\mathrm{J})f^{a_1\ldots a_{k-1}r}.
\label{nablaP}
\end{gather}
Hence all natural operators $\Gamma(S^{k,0}TM) \to \Gamma(S^{k-1,0}TM\otimes F_{2/n})$ are linear
combinations of terms in~\eqref{nabla3}--\eqref{nablaP}.

A similar discussion can be applied to natural operators $\Gamma(S^{k,0}TM) \to \Gamma(S^{k-3,0}TM)$.
In this situation, none of the two indices of ${\bold g}^{-1}$ is contracted in the
expressions~\eqref{terms}.
Reasoning as above, using the properties of symmetry of $\mathrm{C}$ and the fact that $\mathrm{C}$ and $f$
are trace-free, we obtain (since the target bundle is now $S^{k-3,0}TM$) a~simpler list of possible terms:
\begin{gather}
\label{all2}
\nabla_r\nabla_s\nabla_t f^{a_1\ldots a_{k-3}rst},
\qquad
\mathrm{P}_{(rs)_0}\nabla_{t}f^{a_1\ldots a_{k-3}rst},
\qquad
(\nabla_r\mathrm{P}_{st})f^{a_1\ldots a_{k-3}rst},
\end{gather}
for $k \geq 3$.
Hence all natural operators $\Gamma(S^{k,0}TM) \to \Gamma(S^{k-3,0}TM)$ are linear combinations of terms
in~\eqref{all2}.

\vspace{1ex}

II.\ {\bf Conformal invariance.} We shall denote quantities corresponding to the conformally related metric
$\hat{{\rm g}} = e^{2\Upsilon} {\rm g}$ and the corresponding Levi-Civita connection $\hat{\nabla}$ by
$\hat{\mathrm{R}}_{abcd}$, $\hat{\mathrm{P}}_{ab}$, $\hat{\mathrm{J}}$ and~$\hat{\mathrm{A}}_{abc}$.
(The Weyl tensor is missing here since $\hat{\mathrm{C}}_{abcd} = \mathrm{C}_{abcd}$.) This transformation
is controlled by the one-form $\Upsilon_a = \nabla_a \Upsilon$, see e.g.~\cite{BEG94} for details.
Explicitly, one can compute that
\begin{gather}
\label{Rhotrans}
\widehat{\mathrm{P}}_{ab}=\mathrm{P}_{ab}-\nabla_a\Upsilon_b+\Upsilon_a\Upsilon_b-\frac{1}{2}
\Upsilon^r\Upsilon_r{\rm g}_{ab},
\\
\label{JAtrans}
\widehat{\mathrm{J}}=\mathrm{J}-\nabla^r\Upsilon_r-\frac{n-2}{2}\Upsilon^r\Upsilon_r
\qquad
\text{and}
\qquad
\widehat{\mathrm{A}}_{abc}=\mathrm{A}_{abc}+\Upsilon_r\mathrm{C}_{bc}{}^r{}_a
\end{gather}
and also that
\begin{gather}
\label{naRhotrans}
\widehat{\nabla}_{(a}\widehat{\mathrm{P}}_{bc)_0}=\nabla_{(a}\mathrm{P}_{bc)_0}-\nabla_{(a}
\nabla_b\Upsilon_{c)_0}+4\Upsilon_{(a}\nabla_b\Upsilon_{c)_0}-4\Upsilon_{(a}\Upsilon_b\Upsilon_{c)_0}-2\Upsilon_{(a}\mathrm{P}_{bc)_0},
\\
\label{naJtrans}
\widehat{\nabla}_a\widehat{\mathrm{J}}=\nabla_a\mathrm{J}
-\nabla_a\nabla^r\Upsilon_r-(n-2)\Upsilon^r\nabla_r\Upsilon_a+2\Upsilon_a\nabla^r\Upsilon_r-2\Upsilon_a\mathrm{J}+(n-2)\Upsilon_a\Upsilon^r\Upsilon_r,
\\
\widehat{\nabla}_{\big(a}\mathrm{C}_b^{}{}^d{}_{c\big)_0}\!{}^e
=\nabla_{(a}\mathrm{C}_b^{}{}^d{}_{c)_0}\!{}^e-4\Upsilon_{(a}\mathrm{C}_b^{}{}^d{}_{c)_0}\!{}^e
+2\Upsilon_r\delta_{(a}\!{}^{(d}\mathrm{C}_b^{}{}^{e)}{}_{c)}{}^r.
\label{naCtrans}
\end{gather}

We shall start with operators $\Gamma(S^{k,0}TM) \to \Gamma(S^{k-1,0}TM \otimes F_{2/n})$.
First observe that the space of such natural and conformally invariant operators is trivial in the f\/lat
case~\cite{BCo85a,BCo85b} hence the two terms of~\eqref{nabla3} cannot appear.
We need to know how remaining terms in~\eqref{nablaC} and~\eqref{nablaP} transform under the conformal
rescaling $\hat{{\rm g}} = e^{2\Upsilon} {\rm g}$.
First observe that the rescaling of f\/irst order expressions we need is
\begin{gather}
\widehat{\nabla}_r f^{a_1\ldots a_{k-1}r}=\nabla_r f^{a_1\ldots a_{k-1}r}
+(n+2k-2)\Upsilon_r f^{a_1\ldots a_{k-1}r},
\qquad
\notag
\\
\widehat{\nabla}_{[b}f_{c]}{}^{a_1\ldots a_{k-1}}=\nabla_{[b}f_{c]}{}^{a_1\ldots a_{k-1}}
+(k+1)\Upsilon_{[b}f_{c]}{}^{a_1\ldots a_{k-1}}+(k-1)\Upsilon_r^{}\delta_{[b}^{(a_1}f_{c]}^{}{}_{}
^{a_2\ldots a_{k-1})r},\notag
\\
\widehat{\nabla}^{(a_1}f^{a_2\ldots a_{k-1})_0}{}_{bc}
=\nabla^{(a_1}f^{a_2\ldots a_{k-1})_0}{}_{bc}+2\Upsilon^{(a_1}f^{a_2\ldots a_{k-1})_0}{}_{bc}-2\Upsilon_{(b}f_{c)}{}^{a_1\ldots a_{k-1}}
\nonumber
\\
\hphantom{\widehat{\nabla}^{(a_1}f^{a_2\ldots a_{k-1})_0}{}_{bc}=}{}
+2\Upsilon_r^{}\delta_{(b}^{(a_1}f_{c)}^{}{}_{}^{a_2\ldots a_{k-1})_0r},\notag
\\
\widehat{\nabla}_{(b}f_{c)}{}^{a_1\ldots a_{k-1}}=\nabla_{(b}f_{c)}{}^{a_1\ldots a_{k-1}}
+(k-1)\Upsilon_{(b}f_{c)}{}^{a_1\ldots a_{k-1}}-(k-1)\Upsilon^{(a_1}f^{a_2\ldots a_{k-1})}{}_{bc}\notag
\\
\hphantom{\widehat{\nabla}_{(b}f_{c)}{}^{a_1\ldots a_{k-1}}=}{}
+{\rm g}_{bc}\Upsilon_r f^{a_1\ldots a_{k-1}r}+(k-1)\Upsilon_r^{}\delta_{(b}^{(a_1}f_{c)}^{}{}_{}
^{a_2\ldots a_{k-1})r}.
\label{naf}
\end{gather}

We are interested in linear combinations of terms in~\eqref{nablaC} and~\eqref{nablaP} which are
independent on the rescaling $\hat{{\rm g}} = e^{2\Upsilon} {\rm g}$.
Considering formulas~\eqref{Rhotrans}--\eqref{naf}, we observe that the term $\nabla_{(a} \nabla_b
\Upsilon_{c)_0}$ appears only in~\eqref{naRhotrans} and the term $\nabla_a \nabla^r\Upsilon_r$ appears only
on the right hand side of~\eqref{naJtrans}.
This means, terms $(\nabla _r \mathrm{J}) f^{a_1 \ldots a_{k-1}r}$ and $(\nabla_{(r} \mathrm{P}_{st)_0} )
{\bold g}^{t(a_1} f^{a_2 \ldots a_{k-1})_0 rs}$ do not appear in the required linear combination.

The Weyl tensor appears in the conformal transformation of the terms in~\eqref{nablaC} but not of the ones
in~\eqref{nablaP}.
Therefore, we look for conformally invariant linear combinations
\begin{gather}
x_1\mathrm{C}^{r}{}_{st}{}^{(a_1}\nabla_r f^{a_2\ldots a_{k-1})_0st}+x_2
\mathrm{C}^{(a_1}{}_r{}^{a_2}{}_s\nabla_t f^{a_3\ldots a_{k-1})_0rst}
\nonumber
\\
\qquad{}
+x_3\big(\nabla_r\mathrm{C}_s{}^{(a_1}{}_t{}^{a_2})f^{a_3\ldots a_{k-1}\big)_0rst}+x_4
\mathrm{A}_{rs}{}^{(a_1}f^{a_2\ldots a_{k-1})_0rs}
\label{Weyl}
\end{gather}
and{\samepage
\begin{gather}
 y_1\mathrm{J}\nabla_r f^{a_1\ldots a_{k-1}r}+y_2\overline{\mathrm{P}}_{rs}\nabla^{(r}f^{s)a_1\ldots a_{k-1}}
\nonumber
\\
 \qquad{}
+y_3\overline{\mathrm{P}}_r{}^{(a_1}\nabla_s f^{a_2\ldots a_{k-1})_0rs}+y_4
\overline{\mathrm{P}}_{rs}\nabla^{(a_1}f^{a_2\ldots a_{k-1})_0rs},
\label{notWeyl}
\end{gather}
where $\overline{\mathrm{P}}_{rs} = \mathrm{P}_{(rs)_0}$ denotes the trace-free part of $\mathrm{P}$.
In other words, we search for scalars $x_i, y_j \in \mathbb{R}$ such that both~\eqref{Weyl}
and~\eqref{notWeyl} are invariant independently.}

First we discuss~\eqref{Weyl} which is possible only for $k \geq 2$ and some terms only for $k \geq 3$.
Assuming $k \geq 3$, conformal transformations of these terms are
\begin{gather*}
\mathrm{C}^r{}_{st}{}^{(a_1}\widehat{\nabla}_r f^{a_2\ldots a_{k-1})_0st}=\mathrm{C}^{r}{}_{st}{}
^{(a_1}\widehat{\nabla}_r f^{a_2\ldots a_{k-1})_0st}+(k+1)\Upsilon_r\mathrm{C}^{r}{}_{st}{}^{(a_1}
f^{a_2\ldots a_{k-1})_0st}
\\
\hphantom{\mathrm{C}^r{}_{st}{}^{(a_1}\widehat{\nabla}_r f^{a_2\ldots a_{k-1})_0st}=}{}
+(k-2)\mathrm{C}^{(a_1}{}_{st}{}^{a_2}\Upsilon_r f^{a_3\ldots a_{k-1})_0str},
\\
\mathrm{C}^{(a_1}{}_r{}^{a_2}{}_s\widehat{\nabla}_t f^{a_3\ldots a_{k-1})_0rst}=\mathrm{C}^{(a_1}{}
_r{}^{a_2}{}_s\nabla_t f^{a_3\ldots a_{k-1})_0rst}-(n+2k-2)C^{(a_1}{}_{rs}{}^{a_2}
\Upsilon_t f^{a_3\ldots a_{k-1})_0rst},
\\
\big(\widehat{\nabla}_r\mathrm{C}_s{}^{(a_1}{}_t{}^{a_2})f^{a_3\ldots a_{k-1}\big)_0rst}
=(\nabla_r\mathrm{C}_s{}^{(a_1}{}_t{}^{a_2})f^{a_3\ldots a_{k-1})_0rst}+4\mathrm{C}^{(a_1}{}_{rs}{}^{a_2}
\Upsilon_t f^{a_3\ldots a_{k-1})_0rst}
\\
\hphantom{\big(\widehat{\nabla}_r\mathrm{C}_s{}^{(a_1}{}_t{}^{a_2})f^{a_3\ldots a_{k-1}\big)_0rst}=}{}
-2\Upsilon_r\mathrm{C}^{r}{}_{st}{}^{(a_1}f^{a_2\ldots a_{k-1})_0st},
\\
\widehat{\mathrm{A}}_{rs}{}^{(a_1}f^{a_2\ldots a_{k-1})_0rs}=\mathrm{A}_{rs}{}^{(a_1}
f^{a_2\ldots a_{k-1})_0rs}+\Upsilon_r\mathrm{C}^{r}{}_{st}{}^{(a_1}f^{a_2\ldots a_{k-1})_0st}
\end{gather*}
using~\eqref{naf},~\eqref{naCtrans} and~\eqref{JAtrans}.
Now, considering where the term $\Upsilon_r \mathrm{C}^{r}{}_{st}{}^{(a_1} f^{a_2 \ldots a_{k-1})_0st}$
appears in the previous display, we see that $(k+1)x_1 -2x_3 + x_4=0$.
Considering the other term $\mathrm{C}^{(a_1}{}_{rs}{}^{a_2} \Upsilon_t f^{a_3 \ldots a_{k-1})_0rst}$, we
conclude that $(k-2)x_1 - (n+2k-2)x_2 + 4x_3 =0$.
Solutions of this pair of linear equations are generated by $(x_1,x_2,x_3,x_4) = \bigl( n+2k-2,k-2,0,
-(k+1)(n+2k-2) \bigr)$ and $(x_1,x_2,x_3,x_4) = \bigl(0,4, n+2k-2,2(n+2k-2) \bigr)$, therefore the space of
corresponding invariant linear operators is generated by the operators $\mathbf{F}_{1}$ and
$\mathbf{F}_{2}$ def\/ined in the following way:
\begin{gather*}
(\mathbf{F}_{1}(f))^{a_1\ldots a_{k-1}}=\mathrm{C}^{r}{}_{st}{}^{(a_1}\nabla_r f^{a_2\ldots a_{k-1}
)_0st}-(k+1)\mathrm{A}_{rs}{}^{(a_1}f^{a_2\ldots a_{k-1})_0rs}
\\
\hphantom{(\mathbf{F}_{1}(f))^{a_1\ldots a_{k-1}}=}{}
+\frac{k-2}{n+2k-2}\mathrm{C}^{(a_1}{}_r{}^{a_2}{}_s\nabla_t f^{a_3\ldots a_{k-1})_0rst},
\\
(\mathbf{F}_{2}(f))^{a_1\ldots a_{k-1}}=4\mathrm{C}^{(a_1}{}_r{}^{a_2}{}
_s\nabla_t f^{a_3\ldots a_{k-1})_0rst}+(n+2k-2)(\nabla_r\mathrm{C}_s{}^{(a_1}{}_t{}^{a_2})f^{a_3\ldots a_{k-1})_0rst}
\\
\hphantom{(\mathbf{F}_{2}(f))^{a_1\ldots a_{k-1}}=}{}
+2(n+2k-2)\mathrm{A}_{rs}{}^{(a_1}f^{a_2\ldots a_{k-1})_0rs}.
\end{gather*}
This shows that the operators in the statement of the proposition for $k \geq 3$ are invariant.

In the case $k=2$ only some terms from~\eqref{Weyl} can appear.
Specif\/ically, we study the conformal invariance of the linear combination
\begin{gather}
\label{F}
x_1\mathrm{C}^{r}{}_{st}{}^{a}\nabla_r f^{st}+x_4\mathrm{A}_{rs}{}^{a}f^{rs}
\end{gather}
for the section $f^{ab}$ in $\Gamma(S^{2,0}TM)$.
Since $\widehat{\nabla}_{a} f^{bc} = \nabla_{a} f^{bc} + 2 \Upsilon_{a} f^{bc} - 2 \Upsilon^{(b}
f^{c)}{}_{a} + 2\delta_a\!{}^{(b} \Upsilon_r f^{c)}{}^r$, conformal transformations of terms in the
previous display are
\begin{gather*}
\mathrm{C}^r{}_{st}{}^{a}\widehat{\nabla}_r f^{st}=\mathrm{C}^{r}{}_{st}{}^{a}\widehat{\nabla}
_r f^{st}+3\Upsilon_r\mathrm{C}^{r}{}_{st}{}^{a}f^{st}
\qquad
\text{and}
\\
\widehat{\mathrm{A}}_{rs}{}^{a}f^{rs}=\mathrm{A}_{rs}{}^{a}f^{rs}+\Upsilon_r\mathrm{C}^{r}{}_{st}{}
^{a}f^{st}.
\end{gather*}
By the same reasoning as in the case $k \geq 3$, we obtain that the operator given in~\eqref{F} is
invariant if and only if $(x_{1},x_{4})$ is a~multiple of $(1,-3)$.
In the case $k=2$, the only invariant operators are thus the multiples of the operator $\mathbf{F}$
def\/ined by
\begin{gather*}
(\mathbf{F}(f))^a=\mathrm{C}^{r}{}_{st}{}^{a}\nabla_r f^{st}-3\mathrm{A}_{rs}{}^{a}f^{rs}.
\end{gather*}

Now we shall discuss terms~\eqref{notWeyl} and we assume $k \geq 2$ f\/irst.
Consider an arbitrary but f\/ixed point $x \in M$.
We can choose the function $\Upsilon$ such that $\Upsilon_a(x)=0$, $\nabla_{(a}\Upsilon_{b)_0}(x) =
\Phi_{ab}(x)$ and $\nabla^r\Upsilon_r(x) = \Psi(x)$ for any prescribed values of $\Phi_{ab}(x)$ and
$\Psi(x)$.
Therefore, the conformal transformation of terms in~\eqref{notWeyl} is
\begin{gather*}
\widehat{\mathrm{J}}\widehat{\nabla}_r f^{a_1\ldots a_{k-1}r}=\mathrm{J}\nabla_r f^{a_1\ldots a_{k-1}
r}-\Psi\nabla_r f^{a_1\ldots a_{k-1}r},
\\
\widehat{\overline{\mathrm{P}}}_{rs}\widehat{\nabla}^{(r}f^{s)a_1\ldots a_{k-1}}=\overline{\mathrm{P}
}_{rs}\nabla^{(r}f^{s)a_1\ldots a_{k-1}}-\Phi_{rs}\nabla^{(r}f^{s)a_1\ldots a_{k-1}},
\\
\widehat{\overline{\mathrm{P}}}_{r}{}^{(a_1}\widehat{\nabla}_s f^{a_2\ldots a_{k-1})_0rs}
=\overline{\mathrm{P}}_{r}{}^{(a_1}\nabla_s f^{a_2\ldots a_{k-1})_0rs}-\Phi_{r}{}^{(a_1}
\nabla_s f^{a_2\ldots a_{k-1})_0rs},
\\
\widehat{\overline{\mathrm{P}}}_{rs}\widehat{\nabla}^{(a_1}f^{a_2\ldots a_{k-1})_0rs}
=\overline{\mathrm{P}}_{rs}\nabla^{(a_1}f^{a_2\ldots a_{k-1})_0rs}-\Phi_{rs}\nabla^{(a_1}
f^{a_2\ldots a_{k-1})_0rs}
\end{gather*}
at the point $x$ (which is for simplicity omitted in the previous display).
Choosing $\Psi(x) \not =0$ and $\Phi_{rs}(x)=0$, the invariance of~\eqref{notWeyl} means that $y_1=0$.
Henceforth we assume $\Psi(x)=0$ and $\Phi_{rs}(x)\not=0$.
To determine $y_2$, $y_3$ and $y_4$, we shall test invariance of~\eqref{notWeyl} for $f^{a_1 \ldots a_k}$
with specif\/ic properties at $x$.
First assume that $\nabla^b f^{a_1 \ldots a_k}(x) = \nabla^{(b} f^{a_1 \ldots a_k)}(x)$, or equivalently
that $\nabla^b f^{a_1 \ldots a_k}(x) = \nabla^{a_1} f^{ba_2 \ldots a_k}(x)$.
This in particular implies that $\nabla_r f^{a_1 \ldots a_{k-1}r}(x)=0$ and the invariance
of~\eqref{notWeyl} then means that $y_2+y_4=0$.
Second, we assume $\nabla^{(b} f^{a_1 \ldots a_k)}(x) =0$ or equivalently $2\nabla^{(b} f^{c)a_1 \ldots
a_{k-1}}(x) + (k-1) \nabla^{(a_1} f^{a_2 \ldots a_{k-1})bc}(x) =0$.
This also implies $\nabla_r f^{a_1 \ldots a_{k-1}r}(x)=0$ and the invariance of~\eqref{notWeyl} now means
that $- \frac{k-1}{2} y_2+y_4=0$.
Overall, this yields $y_2 = y_4 =0$, and $y_3=0$ follows.
All scalars in~\eqref{notWeyl} are thus equal to zero.
This completes the proof of the part $(ii)$ of the proposition.

If $k=1$,~\eqref{notWeyl} reduces to the linear combination $y_1  \mathrm{J} \nabla_r f^{ra} + y_2
\overline{\mathrm{P}}_{rs} \nabla^{r} f^{sa}$.
As above, the choice $\Psi(x) \not=0$ and $\Phi_{rs}(x)=0$ shows that $y_1=0$.
Hence $y_2=0$ and the part $(i)$ follows.

In order to complete the proof of the part $(iii)$, it remains to describe natural and conformally invariant
operators $\Gamma(S^{k,0}TM) \to \Gamma(S^{k-3}TM)$.
The space of these operators is also trivial in the f\/lat case~\cite{BCo85a,BCo85b}, hence the f\/irst term
in~\eqref{all2} cannot appear.
Thus the required operator is a~linear combination of the form
\begin{gather*}
x_1\mathrm{P}_{(rs)_0}\nabla_{t}f^{a_1\ldots a_{k-3}rst}+x_2(\nabla_r\mathrm{P}_{st})f^{a_1\ldots a_{k-3} rst},
\end{gather*}
where $x_1,y_1 \in \mathbb{R}$.
Reasoning similarly as above, we observe that $\nabla_{(a} \nabla_b \Upsilon_{c)_0}$ appears only in the
conformal transformation of the second term in the previous display.
Therefore $x_2=0$, hence also $x_1=0$ and the proposition follows.
\end{proof}

\section[Classif\/ication of second order symmetries of $\Delta_{Y}$]{Classif\/ication of second order symmetries of $\boldsymbol{\Delta_{Y}}$}
\label{sym2nd}

We start this section with the def\/inition of the algebra $\mathcal{A}$ of conformal symmetries of the
conformal Laplacian.
Afterwards, we provide our main result: a~complete description of the space~$\mathcal{A}^2$ of second order
conformal symmetries.

\subsection{The algebra of symmetries of the conformal Laplacian}

Let $(M,[{\rm g}])$ be a conformal  manifold of dimension~$n$.
Fixing a~metric ${\rm g}\in [{\rm g}]$, we can regard the conformal Laplacian,
$\Delta_{Y}=\nabla_a{\rm g}^{ab}\nabla_b-\frac{n-2}{4(n-1)}\Sc$, as acting on functions.
The symmetries of~$\Delta_{Y}$ are def\/ined as dif\/ferential operators which commute with~$\Delta_{Y}$.
Hence, they preserve the eigenspaces of~$\Delta_{Y}$.
More generally, conformal symmetries~$D_1$ are def\/ined by the weaker algebraic condition
\begin{gather}
\label{ConfSymDel}
\Delta_{Y}\circ D_1=D_2\circ\Delta_{Y},
\end{gather}
for some dif\/ferential operator $D_2$, so that they only preserve the kernel of $\Delta_{Y}$.
The operator $\Delta_{Y}$ can be considered in equation~\eqref{ConfSymDel} as acting between dif\/ferent
line bundles and in particular as an element of $\mathcal{D}_{\lambda_0,\mu_0}$, where
$\lambda_{0}=\frac{n-2}{2n}$, $\mu_{0}=\frac{n+2}{2n}$.
With this choice, $\Delta_{Y}$ is conformally invariant and the space of conformal symmetries depends only
on the conformal class of the metric ${\rm g}$.
It is stable under linear combinations and compositions.

The operators of the form $P\Delta_{Y}$, i.e.\ in the left ideal generated by $\Delta_{Y}$, are obviously
conformal symmetries.
Since they act trivially on the kernel of $\Delta_{Y}$, they are considered as trivial.
Following~\cite{Eas05,GSi09,Mic11b}, this leads to
\begin{Definition}
Let $(M,[{\rm g}])$ be a~conformal manifold with conformal Laplacian
$\Delta_{Y}\in\mathcal{D}_{\lambda_0,\mu_0}$.
The algebra of conformal symmetries of $\Delta_{Y}$ is def\/ined as
\begin{gather*}
\mathcal{A}:=\{D_1\in\mathcal{D}_{\lambda_{0},\lambda_{0}}\,\vert\,
\exists\, D_2\in\mathcal{D}_{\mu_{0},\mu_{0}}
\;
\text{s.t.}
\;
D_2\circ\Delta_{Y}=\Delta_{Y}\circ D_1\},
\end{gather*}
and the subspace of trivial symmetries as
\begin{gather*}
(\Delta_{Y}):=\{A\Delta_{Y}\,\vert\, A\in\mathcal{D}_{\mu_{0},\lambda_{0}}\}.
\end{gather*}
\end{Definition}
Thus, $\mathcal{A}$ is a~subalgebra of $\mathcal{D}_{\lambda_0,\lambda_0}$ and $(\Delta_{Y})$ is the left
ideal generated by $\Delta_{Y}$ in $\mathcal{D}_{\lambda_{0},\lambda_{0}}$.
The f\/iltration by the order on $\mathcal{D}_{\lambda_0,\lambda_0}$ induces a~f\/iltration on
$\mathcal{A}$ and we denote by
\begin{gather*}
\mathcal{A}^k:=\mathcal{A}\cap\mathcal{D}_{\lambda_{0},\lambda_{0}}^k
\end{gather*}
the algebra of conformal symmetries of order $k$.
Obviously, $\mathcal{A}^0\simeq\mathbb{R}$ is the space of constant functions, identif\/ied with zero order
operators on $\lambda_{0}$-densities.
Moreover, the invariance of $\Delta_{Y}$ under the action of conformal Killing vector f\/ields,
see~\eqref{InvConfDel}, shows that $\mathcal{A}^1$ is the direct sum of~$\mathcal{A}^0$ with the space of
Lie derivatives $L^{\lambda_{0}}_X \in\mathcal{D}_{\lambda_{0},\lambda_{0}}^1$ along conformal Killing
vector f\/ields~$X$.
Since~$\mathcal{A}$ is an algebra, $\mathcal{A}^2$ contains in particular $L^{\lambda_{0}}_X\circ
L^{\lambda_{0}}_Y$ for $X$, $Y$ conformal Killing vector f\/ields.

\subsection{The algebra of symmetries of the null geodesic f\/low}

Let $(M,{\rm g})$ be a pseudo-Rie\-man\-nian manifold and $(x^i,p_i)$ denote a canonical coordinate system on $T^*M$.
The inverse metric ${\rm g}^{-1}$ pertains to $\Gamma(S^2TM)$ and identif\/ies with $H:={\rm g}^{ij}p_ip_j\in\mathcal{S}_0$,
where $\mathcal{S}_0=\mathrm{Pol}(T^*M)\cong\Gamma(STM)$ (see Section~\ref{basic}).
Along the isomorphism $T^*M\cong TM$ provided by the metric, the Hamiltonian f\/low of $H$ corresponds to the geodesic f\/low of ${\rm g}$.

The symmetries of the geodesic f\/low are given by functions $K\in\mathcal{S}_0$ which Poisson commute with~$H$.
They coincide with the symmetric Killing tensors.
The null geodesic f\/low, i.e.\ the geodesic f\/low restricted to the level set $H=0$, depends only on the
conformal class of~${\rm g}$.
It admits additional symmetries, namely all the functions $K\in\mathcal{S}_0$ such that
\begin{gather*}
\{H,K\}\in(H),
\end{gather*}
where $\{\cdot,\cdot\}$ stands for the canonical Poisson bracket on $T^*M$, def\/ined in~\eqref{Poisson},
and $(H)$ for the ideal spanned by $H$ in $\mathcal{S}_0$.
The linearity and Leibniz property of the Poisson bracket ensure that the space of symmetries of the null
geodesic f\/low is a~subalgebra of $\mathcal{S}_0$.
Besides, remark that all the functions in $(H)$ are symmetries which act trivially on the null geodesic
f\/low.
\begin{Definition}
Let $(M,{\rm g})$ be a~pseudo-Rie\-man\-nian manifold and $H\in\mathcal{S}_0$ the function associated to~${\rm g}$.
The algebra of symmetries of the null geodesic f\/low of ${\rm g}$ is given by the following subalgebra of~$\mathcal{S}_0$,
\begin{gather*}
\mathcal{K}:=\{K\in\mathcal{S}_0\,|\,
\{H,K\}\in(H)\}.
\end{gather*}
\end{Definition}

In particular, the algebra $\mathcal{K}$ contains the ideal $(H)$ of trivial symmetries.
It inherits the gradation of $\mathcal{S}_0$ by the degree,
\begin{gather*}
\mathcal{K}^k:=\mathcal{K}\cap\mathcal{S}^k_0.
\end{gather*}
The space $\mathcal{K}^0$ is the space of constant functions on $T^*M$.
The Hamiltonian f\/lows of functions in~$\mathcal{K}^1$ coincide with the Hamiltonian lift to~$T^*M$ of the
conformal Killing vectors on $(M,[{\rm g}])$.
For higher degrees, the elements in~$\mathcal{K}$ are symmetric conformal Killing tensors whose Hamiltonian
f\/lows do not preserve the conf\/iguration manifold~$M$.
They are symmetries of the whole phase space but not of the conf\/iguration manifold and often named hidden
symmetries by physicists.
\begin{Proposition}
The elements $K\in\mathcal{K}^k$ are symmetric conformal Killing $k$-tensors.
They are characterized equivalently as:
\begin{itemize}\itemsep=0pt
\item symmetric tensors of order $k$ s.t.
{}$\nabla_{(a_0}K_{a_1\dots a_k)_0}=0$, \item symbols of degree $k$ satisfying $\{H,K\}\in(H)$, \item
elements of $\mathcal{S}^k_0$ in the kernel of the conformal Killing operator $\mathbf{G}$
$($see~\eqref{ConfInvG} or~\eqref{ConfInv-p}$)$.
\end{itemize}
\end{Proposition}
The proof is both classical and straightforward, we let it to the reader.
The next proposition is essential to determine the algebra $\mathcal{A}$ of conformal symmetries.
\begin{Proposition}
\label{Prop:GrA}
If $D_1\in\mathcal{A}^k$ then $\sigma_k(D_1)\in\mathcal{K}^k$.
Under the identification ${\mathrm{gr}}\mathcal{D}_{\lambda_0,\lambda_0}\cong\mathcal{S}_0$, the
associated graded algebra ${\mathrm{gr}}\mathcal{A}$ becomes a~subalgebra of $\mathcal{K}$ and
${\mathrm{gr}}(\Delta_{Y})$ identifies with~$(H)$.
\end{Proposition}
\begin{proof}
Suppose that $D_1$ is a~conformal symmetry of order $k$, i.e.\ satisf\/ies $\Delta_{Y} \circ
D_1=D_2\circ\Delta_{Y}$ for some $D_2$.
Working in the algebra $\mathcal{D}_{\lambda_0,\lambda_0}$ we deduce that $[\Delta_{Y},D_1]\in(\Delta_{Y})$
and the property~\eqref{SymbolPoisson} leads then to $\{H,\sigma_k(D_1)\}\in(H)$, i.e.\
$\sigma_k(D_1)\in\mathcal{K}^k$.
The inclusion ${\mathrm{gr}}\mathcal{A}\leq \mathcal{K}$ follows.
As $\sigma_2(\Delta_{Y})=H$, the property~\eqref{SymbolProduct} of the principal symbol maps implies that
${\mathrm{gr}}(\Delta_{Y})\cong(H)$.
\end{proof}

\subsection{Second order conformal symmetries } We adapt the strategy used in~\cite{Mic11b}, dealing with
conformally f\/lat manifolds.
Thanks to a~natural and conformally invariant quantization, we get a~f\/irst description of the potential
obstruction for a~conformal Killing tensor giving rise to a~conformal symmetry of $\Delta_{Y}$.
\begin{Theorem}
\label{Thm:QK}
Let $\mathcal{Q}_{\lambda,\mu}$ be a~family of natural and conformally invariant quantizations as in
Theorem~{\rm \ref{Thm:NCIQ}}.
We get then
\begin{gather}
\label{QDelta}
\Delta_{Y}\circ\mathcal{Q}_{\lambda_{0},\lambda_{0}}(S)-\mathcal{Q}_{\mu_{0},\mu_{0}}(S)\circ\Delta_{Y}
=\mathcal{Q}_{\lambda_{0},\mu_{0}}\bigl(2\mathbf{G}(S)+\mathbf{Obs}(S)\bigr),
\qquad
\forall\, S\in\mathcal{S}^{\leq2}_0.
\end{gather}
The operator $\mathbf{Obs}$ is the natural and conformally invariant operator defined by
\begin{gather*}
\mathbf{Obs}=\frac{2(n-2)}{3(n+1)}\mathbf{F},
\end{gather*}
where $(\mathbf{F}(S))^a = \mathrm{C}^{r}{}_{st}{}^{a} \nabla_r S^{st} -3 \mathrm{A}_{rs}{}^{a} S^{rs}$ for
$S\in\mathcal{S}^{2}_0$ and we set $\mathbf{F}(S)=0$ for $S\in\mathcal{S}^{\leq 1}_0$.
\end{Theorem}
\begin{proof}
According to~\eqref{S_ks}, we have $\mathcal{S}^2_0=\mathcal{S}^{2,0}_0\oplus\mathcal{S}^{2,1}_0$ and
$S\in\mathcal{S}^{2,1}_0$ is of the following form $S=(|\mathrm{Vol}_{\rm g}|^{\delta_0}H) S_0$ with
$S_0\in \mathcal{F}_{-\delta_0}$.
By Theorem~\ref{Thm:NCIQ}, we have the identities
\begin{gather*}
\mathcal{Q}_{\lambda_{0},\lambda_{0}}(S)=\mathcal{Q}_{\mu_{0},\lambda_{0}}(S_0)\circ\Delta_{Y}
\qquad
\text{and}
\qquad
\mathcal{Q}_{\mu_{0},\mu_{0}}(S)=\Delta_{Y}\circ\mathcal{Q}_{\mu_{0},\lambda_{0}}(S_0).
\end{gather*}
Besides, from the expressions of the operators $\mathbf{G}$ and $\mathbf{F}$ (see e.g.~\eqref{ConfInv-p}),
we deduce{\samepage
\begin{gather*}
\mathbf{G}(S)=0
\qquad
\text{and}
\qquad
\mathbf{Obs}(S)=0.
\end{gather*}
Hence the equality~\eqref{QDelta} holds for all $S\in\mathcal{S}^{2,1}_0$.}

Next, we def\/ine a~natural and conformally invariant operator $\mathrm{QS}$ on
$\mathcal{D}^2_{\lambda_{0},\lambda_{0}}$ by $D\mapsto\Delta_{Y}\circ D-D\circ\Delta_{Y}$.
Pulling this map back to trace-free symbols via the quantization maps,
\begin{gather*}
\xymatrix{
\mathcal{D}^2_{\lambda_{0},\lambda_{0}}\ar[rr]^{\mathrm{QS}} && \mathcal{D}^3_{\lambda_{0},\mu_{0}} \\
(\mathcal{S}^{\leq 1}_0\oplus\mathcal{S}^{2,0}_0) \ar[u]^{\mathcal{Q}_{\lambda_{0},\lambda_{0}}}\ar[rr]_{\mathrm{CS}}
&& \mathcal{S}^{\leq 3}_{\delta_0}\ar[u]_{\mathcal{Q}_{\lambda_{0},\mu_{0}}}
}
\end{gather*}
this leads to a~natural and conformally invariant operator $\mathrm{CS}$ on $\mathcal{S}^{\leq
1}_0\oplus\mathcal{S}^{2,0}_0$.
Since $\Delta_{Y}$ is formally self-adjoint and the quantization maps satisfy the reality
condition~\eqref{real}, we deduce that, for all $S\in\mathcal{S}^{k,0}_0$,
\begin{gather*}
\Delta_{Y}\circ\mathcal{Q}_{\lambda_{0},\lambda_{0}}(S)-\mathcal{Q}_{\mu_{0},\mu_{0}}(S)\circ\Delta_{Y}
\end{gather*}
is of degree $k+1$ and is formally skew-adjoint (resp.\ self-adjoint) if $k$ is even (resp.\ odd).
As such, it is of the form $\mathcal{Q}_{\lambda_{0},\mu_{0}}(P)$, with
$P\in\mathcal{S}^3_{\delta_0}\oplus\mathcal{S}^1_{\delta_0}$ if $S$ is of degree $2$,
$P\in\mathcal{S}^2_{\delta_0}\oplus\mathcal{S}^0_{\delta_0}$ if $S$ is of degree $1$ and
$P\in\mathcal{S}^1_{\delta_0}$ if $S$ is of degree $0$.
We can reduce accordingly the target space of $\mathrm{CS}$ restricted to homogeneous symbols.
Applying Proposition~\ref{Prop:G} and Proposition~\ref{Prop:Obs}, we deduce that $\mathrm{CS}=a\mathbf{G}+b
\mathbf{F}$ for some real constants $a$, $b$.
We have then
\begin{gather}
\label{aandb}
\Delta_{Y}\circ\mathcal{Q}_{\lambda_{0},\lambda_{0}}(S)-\mathcal{Q}_{\mu_{0},\mu_{0}}(S)\circ\Delta_{Y}
=\mathcal{Q}_{\lambda_{0},\mu_{0}}\bigl(a\mathbf{G}(S)+b\mathbf{F}(S)\bigr),
\qquad
\forall\, S\in\mathcal{S}^{\leq2}_0.
\end{gather}

It is straightforward to prove that $a=2$.
To prove $b=\frac {2(n-2)}{3(n+1)}$, we study a~specif\/ic conformal symmetry of $\Delta_{Y}$.
\begin{Lemma}
Let $\eta$ be the pseudo-Euclidean flat metric of signature $(p,q)$, $h$ a~non-vanishing function on
$\mathbb{R}^2$ and $n=p+q+2$.
Let $(M_{0},{\rm g})$ be the pseudo-Rie\-man\-nian manifold $(\mathbb{R}^{2}\times\mathbb{R}^{n-2},{\rm
g}_{0}\times\eta)$, where the metric on $\mathbb{R}^2$ is determined by $({\rm
g}_{0})^{-1}=h(x_1,x_2)p_1^{2}+p_2^{2}$ in canonical Cartesian coordinates $(x^i,p_i)$ on $T^*\mathbb{R}^n$.
Then, $K=p_{3}^{2}$ is a~Killing tensor on $(M_{0},{\rm g})$, and we have the following relation:
\begin{gather*}
\Delta_{Y}\circ\mathcal{Q}_{\lambda_{0},\lambda_{0}}(K)-\mathcal{Q}_{\mu_{0},\mu_{0}}(K)\circ\Delta_{Y}
=\mathcal{Q}_{\lambda_{0},\mu_{0}}(\mathbf{Obs}(K))\neq0.
\end{gather*}
\end{Lemma}

\begin{proof}
Using the relation that links the coef\/f\/icients of ${\rm g}$ and the Christof\/fel symbols
$\Gamma_{jk}^i$ of the associated Levi-Civita connection, it is obvious that $\Gamma_{jk}^{i}=0$ if at
least one of the indices~$i$, $j$, $k$ is greater than or equal to~$3$.
Thus, the only non-vanishing components of the Riemann tensor and the Ricci tensor associated with ${\rm
g}$ are given by the corresponding components of the Riemann tensor and the Ricci tensor of ${\rm g}_{0}$.
In the same way, the scalar curvature of~${\rm g}$ is equal to the scalar curvature of~${\rm g}_{0}$.

Using these facts and the formula for $\mathcal{Q}_{\lambda_{0},\lambda_{0}}(K)$ presented in the proof of
Proposition~\ref{Prop:QKilling}, it is easy to see that
\begin{gather*}
\mathcal{Q}_{\lambda_{0},\lambda_{0}}(K)=\mathcal{Q}_{\mu_{0},\mu_{0}}(K)=\partial_{x^3}^{2}+\frac{1}
{2(n-1)(n+1)}\Sc.
\end{gather*}
By a~direct computation, we obtain the following relation:
\begin{gather*}
\Delta_{Y}\circ\mathcal{Q}_{\lambda_{0},\lambda_{0}}(K)-\mathcal{Q}_{\mu_{0},\mu_{0}}(K)\circ\Delta_{Y}
\\
\qquad
=[\Delta_{Y},\mathcal{Q}_{\lambda_{0},\lambda_{0}}(K)]=\frac{1}{(n-1)(n+1)}\mathcal{Q}_{\lambda_{0},\mu_{0}}
({\rm g}^{ij}(\partial_{i}\Sc)p_{j})+f,
\end{gather*}
with $f\in\mathcal{C}^{\infty}(M)$.
According to~\eqref{aandb}, the function $f$ vanishes.

Besides, we can compute easily the Cotton--York tensor $\mathrm{A}$ associated with ${\rm g}$.
Indeed, if $\mathrm{P}$ denotes the Schouten tensor, we have
\begin{gather*}
\mathrm{A}_{ijk}=2\nabla_{[i}\mathrm{P}_{j]k}=\frac{2}{n-2}\nabla_{[i}\left(\Ric_{j]k}-\frac{1}{2(n-1)}
{\rm g}_{j]k}\Sc\right).
\end{gather*}
Using the peculiar form of $K$ and the remark done previously about the Christof\/fel symbols and the
curvature tensors of ${\rm g}$, it is obvious that
\begin{gather*}
\mathrm{A}_{ijk}K^{jk}=-\frac{1}{2(n-1)(n-2)}\partial_{i}\Sc
\end{gather*}
for all $i$.
The conclusion follows immediately.
\end{proof}

By naturality of the map $\mathrm{CS}$ def\/ined above, the coef\/f\/icient $b$ in~\eqref{aandb} depends
only on the signature of the metric.
As $b$ is equal to $\frac {2(n-2)}{3(n+1)}$ in the example presented in the previous lemma, where the
dimension $M_{0}$ is of arbitrary dimension $n$ and ${\rm g}$ of arbitrary signature, we conclude that
$b=\frac {2(n-2)}{3(n+1)}$ in~\eqref{aandb}.
\end{proof}

Obviously, we have $\mathbf{Obs}(S)=0$ if $S$ is a~symbol of degree $0$ or $1$.
Thus, we recover that $\mathcal{A}^1\cong\mathcal{K}^1\oplus\mathcal{K}^0$ and the isomorphism is provided
by $\mathcal{Q}_{\lambda_{0},\lambda_{0}}$.
Since the symmetric conformal Killing tensors $K$ satisfy $\mathbf{G} K=0$, we deduce the following
\begin{Corollary}
\label{Cor:QK}
Let $(M,{\rm g})$ be a~pseudo-Rie\-man\-nian manifold of dimension $n$ endowed with a~symmetric conformal
Killing $2$-tensor $K$.
The operator
\begin{gather*}
\mathcal{Q}_{\lambda_{0},\lambda_{0}}(K)=K^{ab}\nabla_a\nabla_b+\frac{n}{n+2}(\nabla_aK^{ab})\nabla_b
\\
\hphantom{\mathcal{Q}_{\lambda_{0},\lambda_{0}}(K)=}{}
+\frac{n(n-2)}{4(n+2)(n+1)}(\nabla_a\nabla_b K^{ab})-\frac{n+2}{4(n+1)}\Ric_{ab}K^{ab},
\end{gather*}
is a~conformal symmetry of $\Delta_{Y}$ if and only if $\mathbf{Obs}(K)=0$.
\end{Corollary}
\begin{proof}
Indeed, the condition is obviously suf\/f\/icient.
Next, the condition is necessary because if $\mathcal{Q}_{\lambda_{0},\lambda_{0}}(K)$ is a~conformal
symmetry of $\Delta_{Y}$, there exists a~dif\/ferential operator $D$ such that
\begin{gather*}
\Delta_{Y}\circ\mathcal{Q}_{\lambda_{0},\lambda_{0}}(K)=D\circ\Delta_{Y}.
\end{gather*}
We have then successively, using Theorem~\ref{Thm:QK}:
\begin{gather*}
0 = \Delta_{Y}\circ\mathcal{Q}_{\lambda_{0},\lambda_{0}}(K)-D\circ\Delta_{Y}
\\
\phantom{0}{} = (\Delta_{Y}\circ\mathcal{Q}_{\lambda_{0},\lambda_{0}}(K)-\mathcal{Q}_{\mu_{0},\mu_{0}}
(K)\circ\Delta_{Y})+(\mathcal{Q}_{\mu_{0},\mu_{0}}(K)\circ\Delta_{Y}-D\circ\Delta_{Y})
\\
\phantom{0}{} = \mathcal{Q}_{\lambda_{0},\mu_{0}}(\mathbf{Obs}(K))+(\mathcal{Q}_{\mu_{0},\mu_{0}}
(K)-D)\Delta_{Y}.
\end{gather*}
The operator $\mathcal{Q}_{\lambda_{0},\mu_{0}}(\mathbf{Obs}(K))$ is of order one but not the operator
$(\mathcal{Q}_{\mu_{0},\mu_{0}}(K)-D)\Delta_{Y}$, unless it vanishes.
Hence, both terms $\mathcal{Q}_{\lambda_{0},\mu_{0}}(\mathbf{Obs}(K))$ and
$(\mathcal{Q}_{\mu_{0},\mu_{0}}(K)-D)\Delta_{Y}$ have to vanish and then $\mathbf{Obs}(K)=0$.
\end{proof}
In particular, on a~conformally f\/lat manifold, all the conformal Killing $2$-tensors give rise to
conformal symmetries of $\Delta_{Y}$ after quantization by $\mathcal{Q}_{\lambda_{0},\lambda_{0}}$, as
proved in~\cite{Mic11b}.
We are now in position to prove our main theorem, which provides a~full description of the conformal
symmetries of $\Delta_{Y}$ given by second order dif\/ferential operators.
The isomorphism $\Gamma(TM)\cong\Gamma(T^*M)$ provided by the metric is denoted by $^\flat$.
{\samepage \begin{Theorem}
\label{ConfSym}
The second order conformal symmetries of $\Delta_{Y}$ are classif\/ied as follows:
\begin{enumerate}[$(i)$]\itemsep=0pt
\item $\mathcal{A}^1=\big\{L_{X}^{\lambda_{0}}+c \,\vert\, c\in\mathbb{R}
\ \text{and} \
X\in\mathcal{K}^1\big\}$,
\item $\mathcal{A}^2/\mathcal{A}^1\cong\{K\in\mathcal{K}^2\, \vert\,
\mathbf{Obs}(K)^{\flat} \text{ is an exact $1$-form}\}$, and if $K\in\mathcal{K}^2$ satisfies
$\mathbf{Obs}(K)^{\flat}$
$=-2df$, with $f\in\mathcal{C}^{\infty}(M)$, the corresponding element in
$\mathcal{A}^2/\mathcal{A}^1$ is given by
\begin{gather*}
\mathcal{Q}_{\lambda_{0},\lambda_{0}}(K)+f.
\end{gather*}
\end{enumerate}
\end{Theorem}}

\begin{proof}
We deduce from Proposition~\ref{Prop:GrA} that the principal symbol $K$ of a~second-order conformal
symmetry $D_1$ is a~symmetric conformal Killing $2$-tensor.
Since quantization maps are bijective, the operator $D_1$ reads as
\begin{gather*}
D_1=\mathcal{Q}_{\lambda_{0},\lambda_{0}}(K+X+f),
\end{gather*}
with $f$ and $X$ symbols of degree $0$ and $1$ respectively.
Theorem~\ref{Thm:QK} implies that
\begin{gather*}
\Delta_{Y}\circ D_1-\mathcal{Q}_{\mu_{0},\mu_{0}}(K+X+f)\circ\Delta_{Y}=\mathcal{Q}_{\lambda_{0},\mu_{0}}
(2\mathbf{G}(X)+\mathbf{Obs}(K)+2\mathbf{G}(f)).
\end{gather*}
Hence $\Delta_{Y} \circ D_1\in (\Delta_{Y})$ leads to $\mathbf{G}( X)\in(H)$.
By def\/inition of $\mathbf{G}$, this means that $\mathbf{G}(X)=0$, i.e.\ $X\in\mathcal{K}^1$.
As the symbols $\mathbf{Obs}(K)$ and $\mathbf{G}(f)$ are of degree $1$, they cannot pertains to
$(\Delta_{Y})$.
Therefore, $\Delta_{Y} \circ D_1\in (\Delta_{Y})$ is equivalent to $X\in\mathcal{K}^1$ and
$\mathbf{Obs}(K)+2\mathbf{G}(f)=0$.

The items $(i)$ and $(ii)$ in the statement of the theorem are then easily proved.
\end{proof}

\subsection{Second order symmetries}

The general formula \eqref{Formula:Q} for the natural and conformally
invariant quantization on symbols of degree $2$ leads to the following result.
\begin{Proposition}
\label{Prop:QKilling}
Let $(M,{\rm g})$ be a~pseudo-Rie\-man\-nian manifold of dimension $n$ endowed with a~symmetric Killing
$2$-tensor $K$.
The operator
\begin{gather}
\label{Formula:QKilling}
\mathcal{Q}_{\lambda_{0},\lambda_{0}}(K)=\mathcal{Q}_{\mu_{0},\mu_{0}}(K) = K^{ab}
\nabla_a\nabla_b+(\nabla_aK^{ab})\nabla_b-\frac{n-2}{4(n+1)}\big(\nabla_a\nabla_b K^{ab}\big)
\\
\nonumber
\phantom{\mathcal{Q}_{\lambda_{0},\lambda_{0}}(K)=}{}
-\frac{n+2}{4(n+1)}\Ric_{ab}K^{ab}+\frac{1}{2(n-1)(n+1)}\Sc\big({\rm g}_{ab}K^{ab}\big),
\end{gather}
is a~symmetry of $\Delta_{Y}$, i.e.\ $[\Delta_{Y},\mathcal{Q}_{\lambda_{0},\lambda_{0}}(K)]=0$, if and only
if $\mathbf{Obs}(K)=0$.
\end{Proposition}
\begin{proof}
Let $(x^i,p_i)$ be a~canonical coordinate system on $T^*M$.
The Killing equation satisf\/ied by $K$ reads as ${\rm g}^{ij}p_i\nabla_jK=0$.
Applying the trace operator $\Tr={\rm g}_{ij}\partial_{p_i}\partial_{p_j}$ we deduce that
\begin{gather*}
{\rm g}^{kl}(\nabla_k\Tr K)\nabla_l=-2(\nabla_iK^{il})\nabla_l
\qquad
\text{and}
\qquad
{\rm g}^{kl}(\nabla_k\nabla_l\Tr K)=-2\nabla_i\nabla_lK^{il}.
\end{gather*}
Moreover, if $\lambda=\mu$ and $\delta=0$, we have $\beta_1-2\beta_2=1$ and
$\beta_3-2\beta_4=\frac{n^2\lambda(1-\lambda)}{(n+1)(n+2)}$, where the $\beta_i$ are def\/ined in~\eqref{beta}.
The formula for the quantization $\mathcal{Q}_{\lambda,\lambda}$ reduces then, for $K$ a~Killing tensor,~to
\begin{gather*}
\mathcal{Q}_{\lambda,\lambda}(K) = K^{ab}\nabla_a\nabla_b+\big(\nabla_aK^{ab}\big)\nabla_b-\frac{n^2\lambda(1-\lambda)}{(n+1)(n+2)}\big(\nabla_a\nabla_b K^{ab}\big)
\\
\phantom{\mathcal{Q}_{\lambda,\lambda}(K) =}{}
-\frac{n^2\lambda(\lambda-1)}{(n-2)(n+1)}\Ric_{ab}K^{ab}+\frac{2n^2\lambda(1-\lambda)}
{(n-2)(n-1)(n+1)(n+2)}\Sc\big({\rm g}_{ab}K^{ab}\big).
\end{gather*}
Since $\lambda_{0}+\mu_{0}=1$ we deduce that
$\mathcal{Q}_{\lambda_{0},\lambda_{0}}(K)=\mathcal{Q}_{\mu_{0},\mu_{0}}(K)$.
In consequence, the equality $[\Delta_{Y},\mathcal{Q}_{\lambda_{0},\lambda_{0}}(K)]=0$ is equivalent to the
fact that $\mathcal{Q}_{\lambda_{0},\lambda_{0}}(K)$ is a~conformal symmetry of $\Delta_{Y}$.
By Corollary~\ref{Cor:QK}, this means that $\mathbf{Obs}(K)=0$.
\end{proof}
As a~straightforward consequence, we get
\begin{Corollary}
Let $(M,{\rm g})$ be a~conformally flat manifold and $K$ be a~Killing $2$-tensor.
Then, we have $[\mathcal{Q}_{\lambda_{0},\lambda_{0}}(K),\Delta_{Y}]=0$.
\end{Corollary}
This corollary enlights some of the results obtained in~\cite{BEHRR11}.
As for conformal symmetries, we provide a~full description of the symmetries of $\Delta_{Y}$ given by
second order dif\/ferential operators.
\begin{Theorem}
\label{Thm:Sym}
The second order symmetries of $\Delta_{Y}$ are exactly the operators
\begin{gather*}
\mathcal{Q}_{\lambda_{0},\lambda_{0}}(K+X)+f,
\end{gather*}
where $X$ is a~Killing vector field, $K$ is a~Killing $2$-tensor such that $\mathbf{Obs}(K)^{\flat}$ is
an exact one-form and $f\in\mathcal{C}^{\infty}(M)$ is defined up to a~constant by
$\mathbf{Obs}(K)^{\flat}=-2df$.
\end{Theorem}
\begin{proof}
Let $D_1$ be a~second order symmetry of $\Delta_{Y}$.
In view of~\eqref{SymbolPoisson}, we can deduce from $[\Delta_{Y},D_1]=0$ that $\{H,\sigma_2(D_1)\}=0$.
This means that $K=\sigma_2(D_1)$ has to be a~symmetric Killing $2$-tensor.
Since quantization maps are bijective, the operator $D_1$ reads as
\begin{gather*}
D_1=\mathcal{Q}_{\lambda_{0},\lambda_{0}}(K+X+f),
\end{gather*}
with $f$ and $X$ symbols of degree $0$ and $1$ respectively.
Theorem~\ref{Thm:QK} implies that
\begin{gather*}
[\Delta_{Y},D_1] = \mathcal{Q}_{\lambda_{0},\mu_{0}}(2\mathbf{G}(X)+\mathbf{Obs}(K)+2\mathbf{G}(f))
\\
\phantom{[\Delta_{Y},D_1] =}{}
+\big(\mathcal{Q}_{\mu_{0},\mu_{0}}(K+X+f)-\mathcal{Q}_{\lambda_{0},\lambda_{0}}(K+X+f)\big)\circ\Delta_{Y}.
\end{gather*}
We have shown that $\mathcal{Q}_{\mu_{0},\mu_{0}}(K)=\mathcal{Q}_{\lambda_{0},\lambda_{0}}(K)$ in
Proposition~\ref{Prop:QKilling}.
Moreover, the general formulas in Theorem~\ref{thm:ExplicitQ} prove that
$\mathcal{Q}_{\mu_{0},\mu_{0}}(f)=\mathcal{Q}_{\lambda_{0},\lambda_{0}}(f)$ and
$\mathcal{Q}_{\mu_{0},\mu_{0}}(X)-\mathcal{Q}_{\lambda_{0},\lambda_{0}}(X)=\frac{2}{n}\nabla_aX^a$.
Hence, we get
\begin{gather*}
[\Delta_{Y},D_1]=\mathcal{Q}_{\lambda_{0},\mu_{0}}(2\mathbf{G}(X))+\frac{2}{n}(\nabla_aX^a)\Delta_{Y}
+\mathcal{Q}_{\lambda_{0},\mu_{0}}(\mathbf{Obs}(K)+2\mathbf{G}(f))
\end{gather*}
and
\begin{gather*}
\sigma_2([\Delta_{Y},D_1])=2\mathbf{G}(X)+\frac{2}{n}(\nabla_aX^a)H.
\end{gather*}
As $S^2TM=S^{2,0}TM\oplus S^{2,1}TM$, each of the two terms in the right hand side of the second equation
are independent.
Therefore, $[\Delta_{Y}, D_1]=0$ is equivalent to $\mathbf{G}(X)=0$, $\nabla_aX^a=0$ and
$\mathbf{Obs}(K)+2\mathbf{G}(f)=0$.
The equations $\mathbf{G}(X)=0$ and $\nabla_aX^a=0$ mean that $X$ is a~conformal Killing vector f\/ield
with vanishing divergence, i.e.\ $X$ is a~Killing vector f\/ield.
Applying the metric, the equation $\mathbf{Obs}(K)+2\mathbf{G}(f)=0$ translates into
$\mathbf{Obs}(K)^{\flat}=-2df$.
The result follows.\looseness=-1
\end{proof}
For comparison, we recall the alternative classif\/ication obtained in~\cite{BCR02'}.
\begin{Theorem}[\cite{BCR02'}] Let $K$ be a~Killing $2$-tensor and put $\mathbf{I}(K)^{ab}=K^{ac}\Ric_{c}^{b}-\Ric^{ac}K_{c}^{b}$.
Then, we have
\begin{gather*}
\big[\nabla_{a}K^{ab}\nabla_{b}+f,\Delta+V\big]=0
\
\Longleftrightarrow
\
K^{ab}(\nabla_a V)-\frac{1}{3}\big(\nabla_b\mathbf{I}(K)^{ab}\big)=\nabla^a f,
\end{gather*}
where $\Delta=\nabla_a{\rm g}^{ab}\nabla_b$ and $f,V\in\mathcal{C}^{\infty}(M)$.
\end{Theorem}
As an advantage of our method, the obtained condition to get a~symmetry (namely $\mathbf{Obs}(K)^\flat$
exact one-form) is conformally invariant and obviously vanishes on conformally f\/lat manifolds.
As an advantage of the approach used in~\cite{BCR02'} and initiated by Carter~\cite{Car77}, one recovers
easily that
\begin{gather*}
[\Delta_{Y},\nabla_aK^{ab}\nabla_b]=0,
\end{gather*}
for all Killing $2$-tensors $K$ on an Einstein manifold.

\subsection{Higher order conformal symmetries}

Up to now we discussed symbols of order $\leq 2$.
The more general version (which we shall state without proof) of Theorem~\ref{Thm:QK} is as follows.
Assume that $\mathcal{Q}_{\lambda,\mu}$ is a~family of natural and conformally invariant quantizations as
in Theorem~\ref{Thm:NCIQ} and let $S$ be a~trace-free symbol $S \in \mathcal{S}^{k,0}_0$.
Then we get
\begin{gather}
\label{higher}
\Delta_{Y}\circ\mathcal{Q}_{\lambda_{0},\lambda_{0}}(S)-\mathcal{Q}_{\mu_{0},\mu_{0}}(S)\circ\Delta_{Y}
=\mathcal{Q}_{\lambda_{0},\mu_{0}}\bigl(2\mathbf{G}(S)+x\mathbf{F}_{1}(S)+y\mathbf{F}_{2}(S)+\Phi(S)\bigr),
\end{gather}
where operators $\mathbf{F}_{1},\mathbf{F}_{2}:
\mathcal{S}^{k,0}_0 \to \mathcal{S}^{k-1,0}_{\delta_0}$ are
def\/ined in Proposition~\ref{Prop:Obs}, scalars $x$ and $y$ have the value
\begin{gather*}
x=\frac{k(k-1)(n+2k-6)}{3(n+2k-2)(n+2k-3)}
\qquad
\text{and}
\qquad
y=\frac{k(k-1)(k-2)(n+2k)}{12(n+2k-2)(n+2k-3)},
\end{gather*}
and $\Phi$ is a~natural and conformally invariant operator $\Phi:
\mathcal{S}^{k,0}_0 \to \mathcal{S}^{\leq k-3}_{\delta_0}$.
For $k \leq 2$ this recovers Theorem~\ref{Thm:QK}, the general case $k \geq 3$ can be shown by a~direct (but tedius) computation.

Using~\eqref{higher} we can formulate a~higher order version of Corollary~\ref{Cor:QK}: If $K \in
\mathcal{K}^k$ is a~conformal Killing $k$-tensor such that the operator
$\mathcal{Q}_{\lambda_{0},\lambda_{0}}(K)$ is a~conformal symmetry of $\Delta_{Y}$, then
$x\mathbf{F}_{1}(K)+y\mathbf{F}_{2}(K)=0$.
Moreover, the same reasoning as in the proof of Theorem~\ref{ConfSym} yields a~higher order analogue of
this theorem, i.e.\
\begin{gather*}
\mathcal{A}^k/\mathcal{A}^{k-1}\subseteq\bigl\{K\in\mathcal{K}^k\,\vert\,
x\mathbf{F}_{1}(K)+y\mathbf{F}_{2}(K)=\mathbf{G}(\overline{K})
\;
\text{for some}
\;
\overline{K}\in\mathcal{S}^{k-1,0}_0\bigr\}.
\end{gather*}

\section{Examples in dimension 3}\label{Section5}

In this section, we consider the space $\mathbb{R}^{3}$ endowed successively with two types of metrics: the
conformal St\"ackel metrics and the Di Pirro metrics.

The conformal St\"ackel metrics are those for which the Hamilton--Jacobi equation
\begin{gather*}
{\rm g}^{ij}(\partial_iW)(\partial_jW)=E
\end{gather*}
admits additive separation in an orthogonal coordinate system for $E=0$ (see~\cite{BKM78} and references
therein).
They are conformally related to the St\"ackel metrics, for which the additive separation of the
Hamilton--Jacobi equation holds for all $E\in\mathbb{R}$.
Moreover, the separating coordinates, called (conformal) St\"ackel coordinates are characterized by two
commuting (conformal) Killing $2$-tensors.

Except for the St\"ackel metrics, every diagonal metric on $\mathbb{R}^3$ admitting a~diagonal Killing
tensor is a~Di Pirro metric ${\rm g}$ (see~\cite[p.~113]{Per90}), whose corresponding Hamiltonian is (see
e.g.~\cite{DVa05})
\begin{gather}
\label{H-DiPirro}
H={\rm g}^{-1}=\frac{1}{2(\gamma(x_1,x_2)+c(x_3))}\left(a(x_1,x_2)p_1^2+b(x_1,x_2)p_2^2+p_3^2\right),
\end{gather}
where $a,b,c$ and $\gamma$ are arbitrary functions and $(x^i,p_i)$ are canonical coordinates on~$T^*\mathbb{R}^3$.

\subsection{An example of second order symmetry}

The Di Pirro metrics def\/ined via equation~\eqref{H-DiPirro} admit diagonal Killing tensors $K$ given by
\begin{gather*}
K=\frac{1}{\gamma(x_1,x_2)+c(x_3)}
\left(c(x_3)a(x_1,x_2)p_1^2+c(x_3)b(x_1,x_2)p_2^2-\gamma(x_1,x_2)p_3^2\right).
\end{gather*}
For generic functions $a$, $b$, $c$ and $\gamma$, the vector space of Killing 2-tensors is generated by $H$ and
$K$.
However, for some choices of functions, this metric can admit other Killing tensors.
For example, if $(r,\theta)$ denote the polar coordinates in the plane with coordinates $(x_1,x_2)$, if the
functions $a$, $b$, $\gamma$ depend only on $r$ and if $a=b$, then the metric is St\"ackel and admits
$p_{\theta}^{2}$ as additional Killing tensor.
\begin{Proposition}
On the space $\mathbb{R}^{3}$, endowed with the metric ${\rm g}$ defined by~\eqref{H-DiPirro}, there
exists a~symmetry $D$ of $\Delta_{Y}$ whose principal symbol is equal to the Killing tensor~$K$.
In terms of the conformally related metric
\begin{gather*}
\hat{{\rm g}}:=\frac{1}{2(\gamma(x_1,x_2)+c(x_3))}{\rm g},
\end{gather*}
this symmetry is given by: $D=\mathcal{Q}_{\lambda_{0},\lambda_{0}}(K)
+\frac{1}{16}(3\widehat{\Ric}_{ab}-\widehat{{\rm Sc}}\hat{{\rm g}}_{ab})K^{ab}$, i.e.\ by:
\begin{gather*}
D=\widehat{\nabla}_aK^{ab}\widehat{\nabla}_b-\frac{1}{16}(\widehat{\nabla}_a\widehat{\nabla}_b K^{ab}
)-\frac{1}{8}\widehat{\Ric}_{ab}K^{ab},
\end{gather*}
where $\widehat{\nabla}$, $\widehat{\Ric}$ and $\widehat{\Sc}$ represent respectively the
Levi-Civita connection, the Ricci tensor and the scalar curvature associated with the metric $\hat{{\rm
g}}$.
\end{Proposition}
\begin{proof}
We use Theorem~\ref{Thm:Sym}.
In order to compute the obstruction $\mathbf{Obs}(K)^{\flat}$, we used a~Mathematica package called
``Riemannian Geometry and Tensor Calculus'', by Bonanos~\cite{Bon12}.

This obstruction turns out to be an exact one-form equal to $d(-\frac {1}{8}(3\widehat{\Ric}_{ab}-\widehat{{\rm Sc}}\hat{{\rm g}}_{ab})K^{ab})$.
The f\/irst expression of the symmetry $D$ follows, the second one is deduced
from~\eqref{Formula:QKilling}, giving $\mathcal{Q}_{\lambda_{0},\lambda_{0}}(K)$.
\end{proof}

\subsection{An example of obstructions to symmetries}

If written in conformal St\"ackel coordinates, the conformal St\"ackel metrics ${\rm g}$ on $\mathbb{R}^3$
admit four possible normal forms, depending on the numbers of ignorable coordinates (see \cite{BKM78}).
A coordinate~$x$ is ignorable if $\partial_{x}$ is a~conformal Killing vector f\/ield of the metric.

Thus, if $x_1$ is an ignorable coordinate, the conformal St\"ackel metrics ${\rm g}$ read as
\begin{gather}
\label{metric}
{\rm g}=Q\left((dx_{1})^{2}+\big(u(x_2)+v(x_3)\big)\big((dx_2)^{2}+(dx_3)^{2}\big)\right),
\end{gather}
where $Q\in\mathcal{C}^{\infty}(\mathbb{R}^{3})$ is the conformal factor and where $u$ and $v$ are
functions depending respectively on the coordinates $x_2$ and $x_3$.
Such metrics admit $\partial_{x_1}$ as conformal Killing vector f\/ield and
\begin{gather}
\label{Killingreduced}
K=(u(x_2)+v(x_3))^{-1}\big(v(x_3)p_{2}^{2}-u(x_2)p_{3}^{2}\big)
\end{gather}
as conformal Killing $2$-tensor.
\begin{Proposition}
On $\mathbb{R}^{3}$, there exist metrics ${\rm g}$ as in~\eqref{metric} whose conformal Laplacian
$\Delta_{Y}$ admits no conformal symmetry with principal symbol $K$.
\end{Proposition}
\begin{proof}
Indeed, the obstruction associated with $K$, $\mathbf{Obs}(K)^{\flat}$, is generally not closed.
Thanks to the Mathematica package ``Riemannian Geometry and Tensor Calculus'', by Bonanos~\cite{Bon12},
we can actually compute that
\begin{gather*}
d\mathbf{Obs}(K)^{\flat}=-\frac{1}{4}\big(\partial_{x_2}^2+\partial_{x_3}^2\big)\partial_{x_2}\partial_{x_3} \log(u(x_2)+v(x_3)) dx_{2}\wedge dx_{3},
\end{gather*}
where the symbol $'$ denotes the derivatives with respect to the coordinates $x_2$ and $x_3$.
This expression does not vanish e.g.\
for the functions $u(x_2)=x_2$ and $v(x_3)=x_3$.

We conclude then using Theorem~\ref{ConfSym}.
\end{proof}

An example of a~metric of the form~\eqref{metric} is provided by the Minkowski metric on $\mathbb{R}^{4}$
reduced along the Killing vector f\/ield
$X=x_3\partial_{t}+t\partial_{x_3}+a(x_1\partial_{x_2}-x_2\partial_{x_1})$, $a\in\mathbb{R}$
(see~\cite{IKK10}).
In the time-like region of $X$ and in appropriate coordinates $(r,\phi,z)$, the reduced metric is equal to
\begin{gather*}
{\rm g}=dr^2+\frac{r^2z^2}{z^2-a^2r^2}d\phi^2+dz^2
\end{gather*}
and admits $\partial_\phi$ as Killing vector f\/ield.
Moreover, after reduction, the Killing tensor $p_{x_1}^2+p_{x_2}^2$ is equal to
\begin{gather*}
K=p_r^2+\frac{1}{r^2}p_\phi^2.
\end{gather*}
Notice that the metric ${\rm g}$ is a~St\"ackel metric with one ignorable coordinate.
Indeed, the metric takes the form~\eqref{metric}, with $Q(r,z)=\frac{r^2z^2}{z^2-a^2r^2}$, $u(r)=1/r^2$ and
$v(z)=-a^2/z^2$, whereas the conformal Killing tensor $K-\frac{z^2}{z^2-a^2r^2}H$ can be written as
in~\eqref{Killingreduced}.
Here, $H={\rm g}^{-1}$ is the metric Hamiltonian.

In this situation, there is no conformal symmetry of $\Delta_{Y}$ with principal symbol $K$ if $a\neq 0$.
Indeed, the one-form $\mathbf{Obs}(K)^{\flat}$ is then non-exact, as shown by Mathematica computations
\begin{gather*}
d\mathbf{Obs}(K)^{\flat}=\frac{3}{2}\big(a+a^3\big)\left(\frac{1}{(z+ar)^4}-\frac{1}{(z-ar)^4}\right)dr\wedge dz.
\end{gather*}
\begin{Remark}
Extending the metric~\eqref{metric} to $\mathbb{R}^n$ as
\begin{gather*}
{\rm g}=Q\left((dx_{1})^{2}+(u(x_2)+v(x_3))\big((dx_2)^{2}+(dx_3)^{2}\big)+(dx^4)^2+\cdots+(dx^n)^2\right),
\end{gather*}
one can check that $K$, given in~\eqref{Killingreduced}, is again a~conformal Killing tensor and that the
one-form $\mathbf{Obs}(K)^\flat$ is in general non-exact.
Thus, there is no conformal symmetry of $\Delta_{Y}$ with principal symbol $K$.
\end{Remark}

\subsection*{Acknowledgements}

A special thanks is due to Jonathan Kress which points out a~mistake in one of the example provided in the
f\/irst version of this paper.
It is a~pleasure to acknowledge also Christian Duval and Galliano Valent for their constant interest in
this work.
Josef \v{S}ilhan would like to thank Pawel Nurowski for helpful discussions.

This research has been partially funded by the Interuniversity Attraction Poles Program initiated by the
Belgian Science Policy Of\/f\/ice.
J.~\v{S}ilhan was supported by the grant agency of the Czech republic under the grant P201/12/G028.

\pdfbookmark[1]{References}{ref}
\LastPageEnding

\end{document}